\documentclass{iopjournal}
\usepackage{amsmath}
\usepackage{amssymb}
\usepackage[square,sort,comma,numbers]{natbib}

\newcommand{\SNR}{\text{SNR}}

\usepackage{xcolor}
\newcommand{\reply}[1]{}

\usepackage{ragged2e}
\AtBeginDocument{\justifying\sloppy}

\begin{document}
\articletype{Paper}
\title{Assessing a Template-Based Approach for Core-Collapse Supernova Gravitational-Wave Detection}

\author{Haakon Andresen$^{1,*}$ , Bella Finkel$^{2}$ \\}
\affil{$^{1}$The Oskar Klein Centre, Department of Astronomy, 
    Stockholm University, AlbaNova, SE-106 91 Stockholm, Sweden \\}
\affil{$^{2}$Department of Mathematics, University of Wisconsin--Madison, Madison, Wisconsin, 53703, USA \\}
\affil{$^*$Author to whom correspondence should be addressed. \\}
\email{haakon.andresen@astro.su.se}

\begin{abstract}
Gravitational waves from core-collapse supernovae are a promising yet challenging target for detection due to the stochastic and complex nature of these signals. Conventional detection methods for core-collapse supernovae rely on excess energy searches because matched filtering has been hindered by the lack of well-defined waveform templates. 
However, numerical simulations of core-collapse supernovae have improved our understanding of the gravitational wave signals they emit, which enables us, for the first time, to
construct a set of templates that closely
resemble predictions from numerical simulations.
In this study, we investigate the possibility of detecting gravitational waves from core-collapse supernovae using template-based methods. We construct a theoretically-informed template bank and use it to recover core-collapse supernova signals injected into real LIGO-Virgo-KAGRA detector data. We consider the signals from three state-of-the-art numerical models, simulated with three different codes. We evaluate the detection efficiency of the template-filtering approach and how well the injected signal is reconstructed. 
For signals whose structure is well captured by our template bank,
we recover $\sim 90\%$ of injections at a distance of 1~kpc and
$\sim 30$--$60\%$ at 2~kpc. In contrast, a model whose signal
differs significantly from the templates is recovered
less efficiently. For many of the recovered events, the underlying
signal characteristics can be reconstructed with an accuracy of
$\sim10-20$\%. We discuss the strengths and limitations of this
approach and identify areas for further improvements for
template-based methods for supernova gravitational-wave
detection. We also present the open-source Python package
\textsc{SynthGrav} used to generate the template bank.
\end{abstract}
\begin{keywords}
{gravitational waves -- supernovae: general -- methods: data analysis}
\end{keywords}



\section{Introduction}
Core-collapse supernovae are among the most energetic events in the universe, producing gravitational waves (GWs) from 
asymmetric mass motions and neutrino emissions during the explosion.
These GW signals offer a unique 
opportunity to probe the physics of core-collapse supernovae and their outcome. It is challenging to detect GWs from core-collapse supernovae because of their typically low amplitudes and highly stochastic nature, 
which are difficult to isolate within the noisy data streams of GW detectors like the LIGO, Virgo, and KAGRA interferometers. As a result, traditional
GW detection efforts for supernovae have relied on excess energy searches, which identify broadband signals without 
specific waveform models, in contrast to the matched-filtering approach that has been successfully applied to binary coalescences
(see for example \cite{KAGRA:2021vkt}).
Matched filtering has traditionally not been used for core-collapse supernovae due to {the assumption that the irregular quality of the signals makes it difficult to construct sufficiently diverse template banks so that the original signal  can be distinguished from detector noise.} Supernova GWs lack the clean structure of binary mergers, instead displaying complex broadband emission. 
Consequently, detection studies primarily rely on 
excess energy methods
\citep{Arnaud_04,Ando_05,Yokozawa_15,Hayama_15,Gossan_16,
Abbott_16,Srivastava_19,Abbott_20,Halim_21,Szczepanczyk_21,
Abbott_21,
Richardson_22,Afle_23,Szczepanczyk_23b,Bruel_23,Szczepanczyk_24,Gill_24},
Bayesian analysis and/or principal component analysis
\citep{Summerscales_08,Powell_16,Powell_17,Powell_18,Roma_19,Afle_21,Powell_22,Raza_22},
and machine learning techniques
\citep{Astone_18,Chan_20,Lopez_21,Mukherjee_21,Edwards_21,Antelis_22,Mitra_23,Casallas-Lagos_23,Abylkairov_24,Abylkairov_24,Eccleston_24,Nunes_24,Morales_25,Wang_26,Akhmetali_26,Sun_26}.

However, recently techniques similar to matched-filtering have seen some
limited application for core-collapse GWs. 
{Drago et al.~}\cite{drago_23} investigated the possibility of exploiting the fact that the standing 
accretion shock instability (SASI) creates similar modulations in both the neutrino \citep{Tamborra_13} 
and the GW signals \citep{Andresen_17}. 
By using the neutrino signal to construct a filter for the GWs, 
\cite{drago_23} reported improved detection efficiencies for nearby 
events over standard excess energy methods. Richardson et al.~\cite{Richardson_24} demonstrated that matched-filtering methods are very promising for the secular GW emission from core-collapse supernovae associated with asymmetric emission of neutrinos and aspherical matter ejection (see, for example, \cite{Epstein_1978,Turner_1978,Muller_97,Richardson_22}). 

Theoretical advances in modelling core-collapse supernovae 
and their associated GW 
signals have provided a clearer understanding of the
expected features of these signals.
Multi-dimensional simulations have revealed {that}
supernova GWs consist of several complex signal components
\citep{Kotake_09,Murphy_09,Marek_09,Scheidegger_10,Yakunin_10,Kotake_11,
muller_e_12,muller_13,cerda-duran_13,Kuroda_14,Yakunin_15,Kuroda_16,
Andresen_17,Kuroda_17,Takiwaki_18,Hayama_18,Morozova_18,OConnor_18,
Radice_19,Andresen_19,Powell_19,Powell_20,Shibagaki_20,Mezzacappa_20,
Zha_20,Vartanyan_20,Andresen_21,Pan_21,Takiwaki_21,Eggenberger-Andersen_21,
Raynaud_22,Vartanyan_22,Jardine_22,Mezzacappa_23,Bugli_23,Vartanyan_23,
Powell_23,Jakobus_23,Pajkos_23,Andresen_24,Choi_24}.
Each component is characterized by a time-dependent central frequency, 
with GWs emitted around this frequency. 
In addition to the main signal components, 
the signals show broadband background emission.

The theoretical understanding of individual emission components has 
improved substantially over the last decade through direct analysis of 
numerical simulations and studies of supernova oscillation modes 
\citep{Murphy_09,muller_13,Fuller_15,Sotani_16,Torres-Forne_18,Morozova_18,
Torres-Forne_19,Torres-Forne_19b,Sotani_19,Sotani_21,Andresen_21,
Rodriguez_23,Wolfe_23,Zha_24}. Matched filtering is optimal for 
known, deterministic signals in stationary Gaussian noise 
\citep{maggiore_07}, but this optimality cannot be assumed to carry over 
to signals with stochastic components. Core-collapse supernova signals 
occupy an intermediate regime: while the precise time-domain realisation 
is stochastic, the overall time-frequency evolution is set by 
deterministic physics. In this work we investigate whether a template 
bank based on the expected time-frequency structure of core-collapse 
signals produces a matched-filter statistic whose distribution under 
signal-plus-noise is sufficiently separated from the noise-only 
distribution to enable detection in a triggered follow-up search.

{Several authors have built on this growing theoretical understanding
to develop detection and characterisation methods that incorporate
prior knowledge of the expected signal morphology \cite{Heng_09,Logue_12,Srivastava_19,Bizouard_21}. 
These approaches
range from principal component analyses of signals from numerical simulations
\citep{Heng_09, Logue_12, Powell_16, Roma_19} and chirplet-based
parametric models used for Bayesian parameter estimation
\citep{Powell_22}, to phenomenological
synthetic-waveform generators \citep{Astone_18, Lopez_21,Casallas-Lagos_23} and
machine-learning classifiers trained either on these phenomenological
waveforms or on simulation waveforms \citep{Chan_20,Iess_20, Mitra_23}.}

This study explores the feasibility of such a template-based approach by 
recovering supernova signals injected into LIGO-Virgo-KAGRA (LVK) 
detector data. We assess the performance of a relatively small template 
bank in terms of detection efficiency and signal reconstruction. Because 
it is not possible to construct a complete template bank that covers all 
possible physical scenarios and all realisations of the underlying 
stochastic physics, we target the part of the signal that is observed in 
the majority of modern numerical simulations: the emission component 
associated with oscillations of the PNS, whose central frequency 
increases secularly with time as the PNS becomes increasingly compact 
\citep{Kotake_09,Murphy_09,Marek_09,muller_e_12,Kuroda_16,Andresen_17,
Kuroda_17,Takiwaki_18,Hayama_18,Morozova_18,OConnor_18,Radice_19,
Andresen_19,Powell_19,Mezzacappa_20,Eggenberger-Andersen_21,Vartanyan_23,
Powell_23}.

{Similarly to previous work, we model the dominant high-frequency
core-collapse supernova emission using synthetic waveforms.
We then use these waveforms
directly as templates in a template-based filter search, similar to how
templates are used in matched filtering searches for compact binary coalescences,
rather than as training data for machine learning or as signal models
for Bayesian inference.}

This paper is structured as follows. In Section~\ref{sec:methods},
we describe the GW signal we inject (Subsection~\ref{sec:gw_signal}),
the template-filtering process (Subsection~\ref{sec:injection_matching}),
and the template bank (Subsections~\ref{sec:templates} and ~\ref{sec:ff} ). We give a detailed description of the Python package \textsc{SynthGrav} used to generate the templates (Subsection~\ref{sec:synthgrav}).
In Section~\ref{sec:optimal_scenario}, we consider an optimal detection scenario to establish a baseline for performance. We present the result of our {search} approach in Section~\ref{sec:detection_reconstruction}, evaluating detection efficiency, signal reconstruction, and related metrics. We briefly discuss false alarms in Section~\ref{sec:false_alarms} and conclude in Section~\ref{sec:conclusions}, where we discuss the implications of our findings and potential avenues for further refinement of supernova GW detection methods.

\section{Methods} \label{sec:methods}
\subsection{Gravitational-Wave Signals} \label{sec:gw_signal}
Computational advances and improved understanding of the physics underlying
core-collapse supernovae have enabled the generation of numerous and progressively more accurate predictions for the GW signals emitted by core-collapse
supernovae (see \cite{Mezzacappa_24} for a review). Recent work (see, for example,
\cite{Murphy_09,muller_13,Andresen_17,Torres-Forne_18,Powell_19}) has also
improved our understanding of the characteristic morphology of core-collapse
supernova GW signals and how their structure is connected to the underlying
physical processes taking place in the core. Most signal predictions now show
qualitatively similar signals exhibiting clear and predictable features 
associated with specific dynamical processes. However, uncertainties 
in the input physics still affect
the accuracy of numerical simulations. Observations of GWs from
a nearby supernova would provide a direct probe of the central engine and help
constrain the physics of the explosion.

We use the GWs predicted by three state-of-the-art numerical models, {each simulated with a different code}. We will refer to the three models as s15fr, s15.01, and D25. For each model, we perform the detection analysis for three different observer directions.

\begin{figure}
    \centering
    \includegraphics[width=1.0\linewidth]{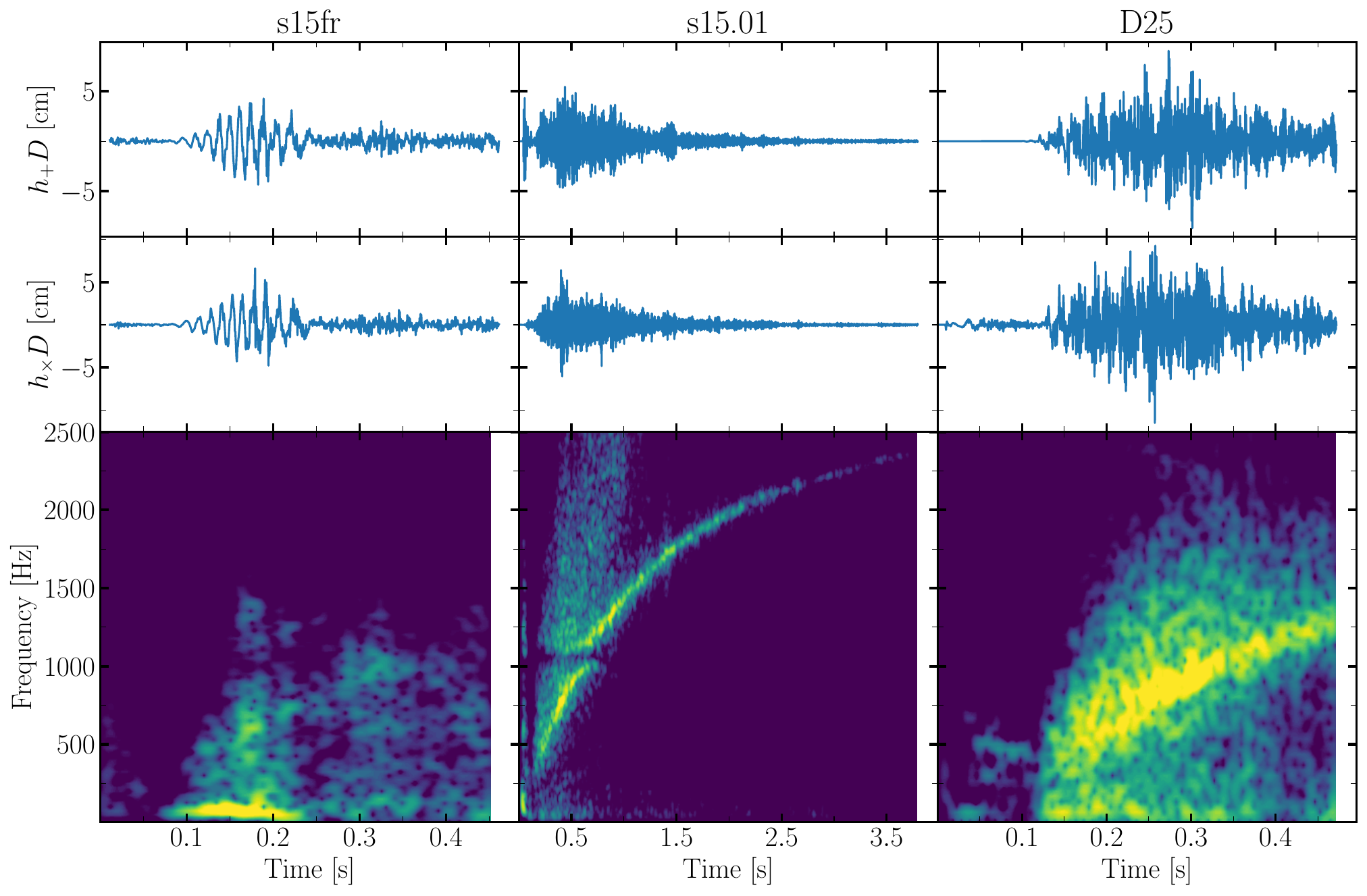}
    \caption{GW signals from the three supernova models considered in this work. The columns correspond to models s15fr (left), s15.01 (middle), and D25 (right). The top rows show the strain components multiplied by the source distance ($D$). The bottom row shows the corresponding spectrograms.}
    \label{fig:signal}
\end{figure}

The GW signals from the models considered in this work are shown in Fig.~\ref{fig:signal} for an observer situated along the $x$-axis of a coordinate system centred on the simulations. Each plot shows the signal from asymmetric matter motions, with the GW strain scaled by the distance to the source.  We plot the two independent components
of the GW tensor in the transverse-traceless gauge. The top panel shows the plus polarisation mode ($h_{+}$) and the middle panel shows the cross polarisation mode ($h_{\times}$). The bottom panels show the spectrograms of the respective signals. We compute the spectrogram by applying short-time Fourier 
transforms (STFT) to $h_{+}$ and $h_{\times}$ individually
before taking the square sum of the two resulting
STFTs. The STFTs are computed with a \texttt{scipy.signal.stft} using a Blackman window \citep{2020SciPy-NMeth}.
We normalize the STFTs and take the base 10 logarithm before plotting.
Time is given in seconds after core bounce.
High-pass and low-pass filters were applied to the signals before
plotting, filtering out the signal below 25 Hz and above 4000 Hz.
We only plot one observer direction in Fig.~\ref{fig:signal} because the
signals for other observer directions look qualitatively similar to the ones shown.

Model s15fr is part of a set of five core-collapse supernova simulations presented in \cite{summa_18}, performed with the \textsc{Prometheus-Vertex} code \cite{rampp_02,Buras_06a,Buras_06b}. 
The GW analysis of a subset of these models was later published in \cite{Andresen_19}. Note that the naming conventions differ between the simulation paper \cite{summa_18} and \cite{Andresen_19}, where the GWs were presented. We adopt the nomenclature in \cite{Andresen_19}.
The simulations were based on a
rotating progenitor with a zero-age main sequence mass of
15 solar masses and with solar metallicity.
Model s15fr was evolved with an artificially enhanced rotation rate (the label ``fr'' denotes fast rotating) relative to the rotation profile obtained from the stellar evolution calculation.
This modification leads to the growth of a strong spiral mode of the SASI. The model underwent successful shock revival around 
0.16 s after bounce.
The resulting GW emission is dominated by a pronounced low-frequency component, producing a narrow-band signal centred around $\sim 75$ Hz. Additionally, the high-frequency signal is weak for model s15fr. The low-frequency signal and the lack of strong high-frequency emission is clearly visible in both the waveforms and the spectrogram of s15fr, see Fig.~\ref{fig:signal}.
This clearly distinguishes the GW 
signal of s15fr from the other two models we considered in our study. 

Model s15.01 is part of a large set of three-dimensional simulations \cite{Vartanyan_23} that were evolved for several seconds with the \textsc{Fornax} code \cite{skinner_19,burrows_20}. The stellar progenitor is a solar metallicity star with an
initial zero-age main sequence mass of 15.01 solar masses. The model
explodes around 0.4 s after bounce. The GW signal from s15.01 is characterised by a narrow emission band at high frequencies which is visible throughout the simulation. The narrow-band emission continues after the onset of shock revival. In addition, emission over a broad range of frequencies is visible prior to $\sim0.6$ post bounce (see the middle row of Fig.~\ref{fig:signal}). The amplitude of the signal is $\sim 5$ cm.
Note that the high-pass filter we have applied to the signal removed a low-frequency component that develops after shock revival. This component is related to the GW memory.

Model D25 was simulated with the \textsc{Chimera} code 
\citep{Bruenn_18}. 
The simulation is based on a non-rotating progenitor with solar metallicity and a zero-age main sequence mass of 25 solar masses \citep{Mezzacappa_23}. We follow the naming convention of the \textsc{Chimera} models
and label the model ``D25'' (D stands for the  D-series of \textsc{Chimera} models). 
Rapid shock expansion sets in at approximately 0.25 s after bounce in the D25 model. The GWs from D25 consist of a strong high-frequency component with amplitudes around $\sim$5 cm and a smaller amount of emission around 100 Hz (see the left column of Fig.~\ref{fig:signal}). As was the case for model s15.01, the high-pass filter we applied to the signals before plotting removed a low-frequency component from the signal from model D25.

\subsection{{Noise Injection and Search Methodology}} \label{sec:injection_matching}
{We inject the signal into an approximately week-long stretch of O3b 
LIGO-Virgo-KAGRA (LVK) data, between GPS times 1266368512 and 
1267063808, obtained from the Gravitational Wave Open Science Center 
(GWOSC) \citep{LVKgwosc}. This interval spans 695{,}296~s 
(approximately 8.05~days). We obtain the data through the GWOSC Python 
interface \citep{LVKgwosc}.
For each detector, the GWOSC strain data is provided in 4096-s chunks. 
Within these chunks, there can be periods of missing data where the strain data
is not available through the GWOSC interface during these times.  
The duration of the missing periods varies, ranging from a small 
fraction of a chunk to most of it.
We split each 4096-s chunk into eight 512-s segments and 
discard any segment that contains missing samples in either the 
Hanford or Livingston detector, since our analysis requires coincident 
two-detector data. After this cut, approximately 5.58~days of usable, 
coincident data remain, corresponding to a joint duty cycle of roughly 
70\%.}

{We consider a two-detector network consisting of the Livingston and Hanford LIGO detectors. We do not include the Virgo detector
in our analysis because 
we do not see a significant
improvement when including three detectors.} We refer the reader to 
\cite{Szczepanczyk_21} for a discussion regarding detector networks
and their importance for the observation of core-collapse supernovae.

{For each of the 512 s long data segments, we inject the signal once into
the data of both detectors. We randomise the sky localisation
of the injections. We use \textsc{GWpy} \citep{gwpy} to
inject the signals into the detector data and to whiten the resulting strain data.
The whitening in \textsc{GWpy} uses inverse spectrum truncation.
We perform {our search }using the functionality of
\textsc{pyCBC} \citep{allen_04,allen_05,dalCanton_14,nitz_17,nitz_24}. We repeat the analysis three times for three
different injection distances: 1 kpc, 2 kpc, and 5 kpc.}

After injecting the signal into the data frames, we
calculate the signal-to-noise ratio (SNR)
using functionality from \textsc{pyCBC}, with
an upper-frequency cut-off of 2000 Hz. 
We use a template bank of 150 templates, see
Section~\ref{sec:templates} for details. 
{When calculating the SNR}, we
estimate the PSD of the noise using the whole 512 s long {segments}. 
We cluster the resulting SNR time series into 1 s long bins and set the SNR of a bin to the
maximum value within that bin. We choose the duration of our bins based on the fact that the
templates are less than half a second long. We calculate the SNR of each template in
both detectors and define the network SNR as
\begin{align}
    \SNR_n^2 = \SNR_L \cdot \SNR_H,
\end{align}
where $\SNR_L$ and $\SNR_H$ represent the SNR values in the Livingston and Hanford
detectors, respectively.
{Our definition of $\SNR_n$ follows \cite{Richardson_24}, {though we note that}
it is more common to define the network SNR in terms of the square sum of the
SNR of each detector.}
{{With our convention}, $\SNR_n$ forms a time series with the same 
1 s long bins as $\SNR_L$ and $\SNR_H$ and we consider a trigger as a bin with an $\SNR_n$ larger than six. The SNR calculation is repeated for each
template. Consequently, each template generates a set of triggers.
For any given injection, the template with the highest SNR is considered the
best match to the injected signal and kept for parameter estimation. 
Importantly, the detection distances considered in this work are well within the Milky Way and we expect that neutrino detectors will pin down the time of the explosion with millisecond precision for a galactic event \citep{Azfar_24}. We therefore assume that the arrival time of the signal is known, allowing false alarms to be discarded if they do not coincide with the neutrino signal. In practice, we disregard triggers that are not consistent with the injection time. {This effectively corresponds to a search window of 1 s.} Accordingly, our analysis is a triggered follow-up search rather than a blind, all-sky search.}
\begin{figure}
    \centering
    \includegraphics[width=1\linewidth]{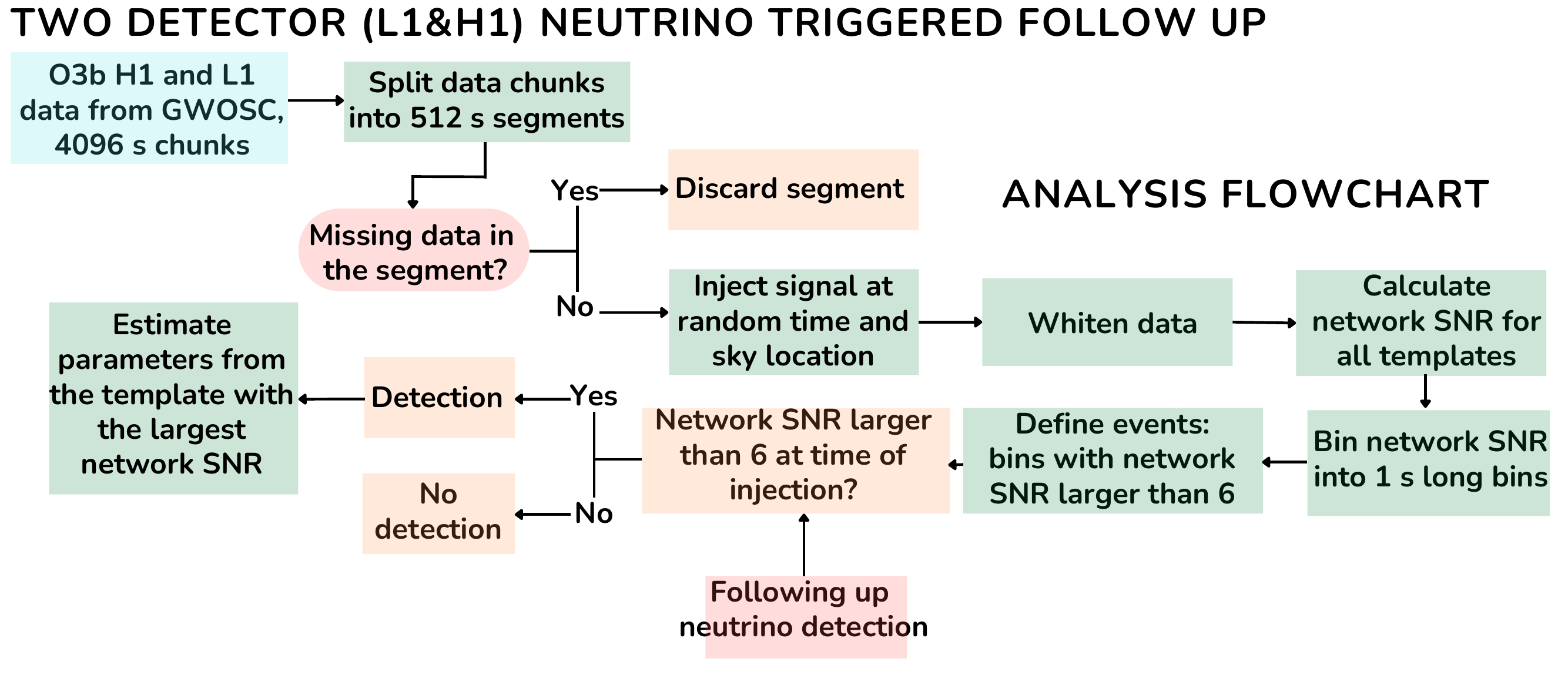}
    \caption{{Flowchart outlining the signal injection and detection procedure. The process includes segmentation of GWOSC data, signal injection, whitening, and event detection based on a network SNR threshold.}}
    \label{fig:flowchart}
\end{figure}

\subsection{Signal Templates} \label{sec:templates}
We generate a set of 150 templates with the newly developed \textsc{SynthGrav}
Python package, see Section \ref{sec:synthgrav}. In short, \textsc{SynthGrav} constructs time-domain signals from one or more signal components. For each component, the user specifies a central frequency as a function of time. The code then generates bursts of
coloured noise that matches the prescribed time-frequency evolution of the signal. The code
does not set the overall amplitude of the templates, but instead normalises the resulting time
series such that it lies between -1 and 1.
As seen in Fig.~\ref{fig:signal}, most of the GW emission from a core-collapse supernova is 
expected to be emitted within 0.2-1.0 seconds after core bounce. Furthermore, the central
frequency of the high-frequency component of the signal is expected to increase almost linearly with time
prior to 1 s post bounce \cite{Andresen_17}. We therefore construct our templates from a single component
with a central frequency of the form
\begin{equation}
    f_c(t) = a t + b.
\end{equation}
We choose three different $b$ values for the sample: 100 Hz, 200 Hz, and 300 Hz. The more
interesting parameter, the slope of the frequency evolution, is sampled in 50 equidistant steps from 250 Hz/s to 3000 Hz/s. For simplicity, we will omit the
units of $a$ and $b$ going forward.
We selected the functional form of $f_c(t)$, 
and the parameters $a$ and $b$, based on well established predictions for the typical
GW emission from core-collapse supernovae \citep{Murphy_09,Marek_09,Kotake_11,
muller_e_12,muller_13,Kuroda_16,
Andresen_17,Takiwaki_18,Hayama_18,Morozova_18,OConnor_18,
Radice_19,Andresen_19,Powell_20,Vartanyan_20,
Takiwaki_21,Eggenberger-Andersen_21,
Jakobus_23}.
In addition to the main signal component, we add a small
amount ($ \lesssim 10\%$ of the total amplitude) of
random noise to mimic the broadband background emission 
seen in theoretical signal predictions.

The way we generate templates does not include any information about the signal amplitude as a function of time, but it can and should be included in the template generation. In principle, one should add the time evolution of the amplitude and the duration of the templates as free
parameters. Doing so, however, increases the size of the template bank and
the computational cost of our analysis. We therefore 
only impose an initial quiescent phase of 0.1 s and construct templates that are $\sim$ 0.45 s long. The templates are long enough to capture the main emission phase, but significantly shorter than the signals from model s15.01. Our aim is to construct the templates with minimal assumptions and ensure they are generic and not tuned to a specific model. This does, however, mean that we are likely to decrease the
detectability of longer signals.

\begin{figure}
    \centering
    \includegraphics[width=1.0\linewidth]{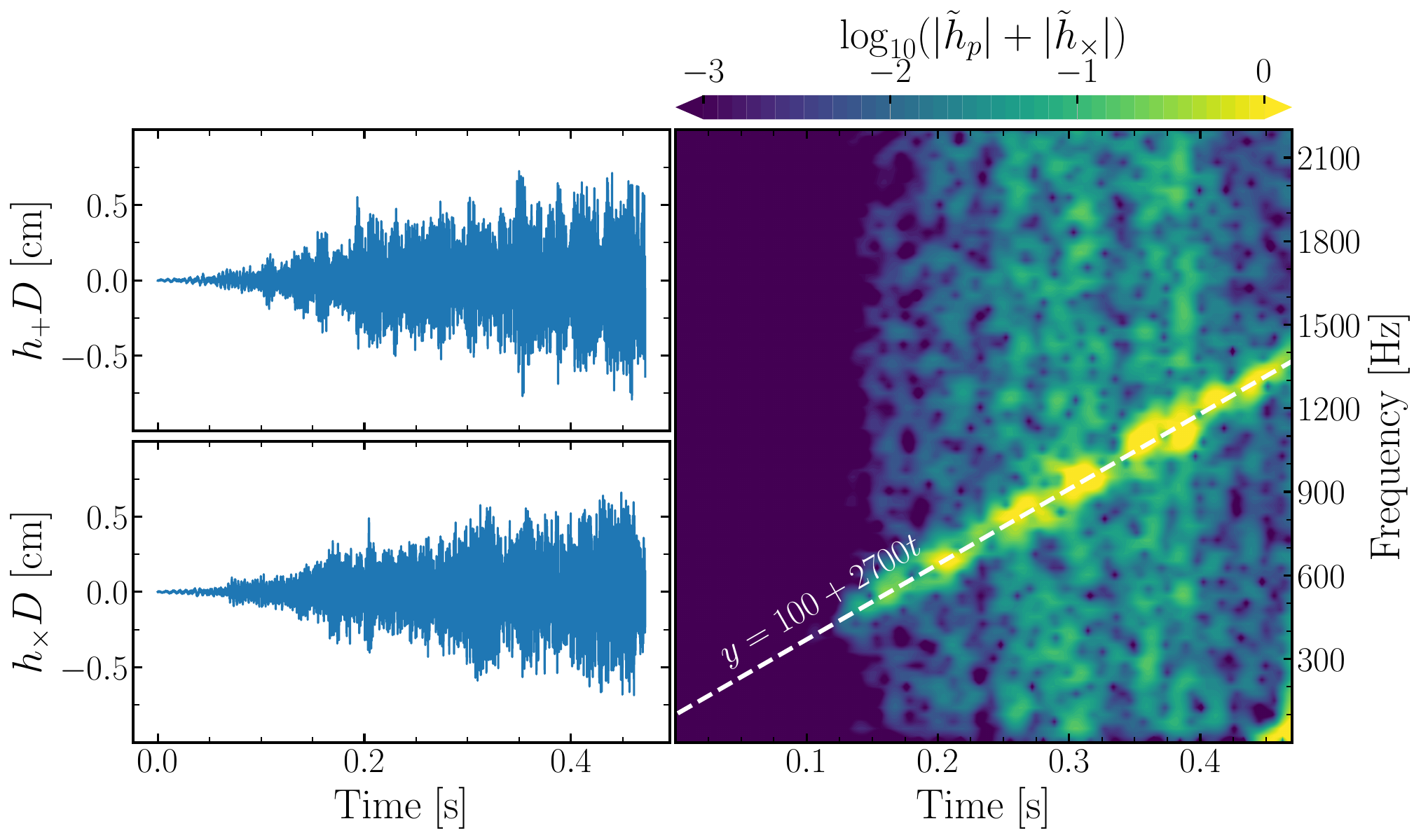}
    \caption{The left panels show an example template in the time domain. The top panel displays the plus polarisation mode and the bottom panel the cross polarisation mode. In both cases, the strain has been multiplied by the distance ($D$) to the source. 
    The right panel shows the corresponding spectrogram of the template. The white dashed line shows the prescribed central frequency evolution used in the template construction for this specific waveform template. The colour bar indicates the logarithmic amplitude scale, after normalising the spectral amplitudes.}
    \label{fig:template}
\end{figure}
Fig.~\ref{fig:template} shows an example template generated with
\textsc{SynthGrav}. The left panels display the $h_+$ and $h_\times$
components of the template, while the right panel shows the
corresponding spectrogram. The dashed white line indicates the central
frequency evolution used to construct the template, given by
$f_c(t) = 100 + 2700t$. The template shows the typical structure of 
a core-collapse GW signal: a high-frequency component
whose central frequency increases approximately linearly with time
after core bounce. While the synthetic template does not reproduce the
full complexity of the simulated signals, it mimics their overall
time–frequency structure of typical signals, compare Fig.~\ref{fig:signal} and
Fig.~\ref{fig:template}.

\subsection{Fitting Factor of the Template Bank}
\label{sec:ff}
{The fitting factor (FF) introduced by \cite{Apostolatos_95} quantifies how
well the best template in a finite bank reproduces a target waveform. We use
it here as a quantitative measure of how well our 150-template bank captures
the morphology of each of the three injected signals.

We define the
noise-weighted inner product between two real time series $a(t)$ and $b(t)$ as
\begin{equation}
\langle a | b \rangle = 4 \mathrm{Re} \int_{f_\mathrm{low}}^{f_\mathrm{high}}
\frac{\tilde{a}(f)  \tilde{b}^*(f)}{S_n(f)} \, df,
\label{eq:inner_product}
\end{equation}
where $\tilde{a}(f)$ is the Fourier transform of $a(t)$, $S_n(f)$ is the
one-sided noise power spectral density (PSD) of the detector, and the integral
is restricted to the analysis band $[20, 2000]\,\mathrm{Hz}$. The normalised match between two waveforms is
\begin{equation}
m(a, b) = \max_{t_0, \phi_0}
\frac{\langle a | b \rangle}{\sqrt{\langle a | a \rangle \, \langle b | b \rangle}},
\label{eq:match}
\end{equation}
maximised over an arbitrary time shift $t_0$ and constant phase $\phi_0$ between
the two waveforms. The fitting factor of the template bank against a target
signal $s(t)$ is the maximum match over the bank,
\begin{equation}
\mathrm{FF}(s) = \max_{k = 1, \ldots, 150}\, m(s, h_k),
\label{eq:ff}
\end{equation}
where $h_k(t)$ is the $k$-th template. By construction, $\mathrm{FF} \in [0, 1]$.

We calculate the fitting factors in the following way:
for each numerical model and each observer direction considered in this work,
we construct the target signal as $s(t) = h_+(t) + h_\times(t)$. The templates are
similarly projected as $h_k(t) = h_{+,k}(t) + h_{\times,k}(t)$. 
We estimate $S_n(f)$ from O3b LIGO data using a $512$-s segment from the
strain frames starting at GPS time $1266966528$, obtained from GWOSC and
processed separately for the Hanford and Livingston detectors. The PSD is
computed using \textsc{pyCBC}.

\begin{table}
\caption{Fitting factors of the 150-template bank against the three 
numerical models considered in this work, computed for the combined 
polarisation $s = h_+ + h_\times$ and for both H1 and L1 PSDs estimated 
from O3b data. Ranges span the three observer directions per model.
\label{tab:fitting_factor}}
\centering
\begin{tabular}{lcc}
\hline \hline
Model    & FF range (H1) & FF range (L1) \\
\hline
D25      & 0.38 -- 0.45  & 0.38 -- 0.46  \\
s15.01   & 0.31 -- 0.40  & 0.30 -- 0.41  \\
s15fr    & 0.08 -- 0.20  & 0.09 -- 0.20  \\
\end{tabular}
\end{table}
The resulting fitting factors are summarised in Table~\ref{tab:fitting_factor}. We find $\mathrm{FF} \approx 0.38-0.46$ 
for the D25 model, $\mathrm{FF} \approx 0.30-0.41$ for s15.01, and 
the substantially lower $\mathrm{FF} \approx 0.08-0.20$ for s15fr. 
The per-detector values for H1 and L1 agree to within a few percent in all 
cases. We found similar numbers when calculating $S_n(f)$ from
the design sensitivity of the two LIGO detectors.
The values for the fitting factors we obtain indicate that our template bank is far
from complete, as should be expected based on its simple construction and
relatively small size.}

{To verify that our choice of 150 templates is reasonable, we repeated the
analysis with reduced banks of 30 and 50 templates. 
Going from 50 to 150 templates changes the number of
detected injections at $1$~kpc by less than $\sim 1\%$ for D25 and by
$\sim 6\%$ for s15.01. For s15fr the gain is larger, $\sim 15\%$, but
the detection efficiency for this model is fundamentally limited by the
morphological mismatch between its signal and the templates we consider here,
as reflected in the low fitting factors (see Table~\ref{tab:fitting_factor}).
The false alarm rate also exhibits only a weak dependence on bank size
(see Section~\ref{sec:false_alarms}). Together, these tests suggest that 150
templates are sufficient to cover the two-dimensional parameter space
$(a, b)$ for the signal morphologies considered in this work. We note
that this contrasts with template banks for compact binary coalescences,
which target a higher-dimensional intrinsic parameter space and require
$\sim 10^5$ templates to achieve $\mathrm{FF} \gtrsim 0.97$
(see, for example, \cite{Owen_99}).}

\subsection{SynthGrav} \label{sec:synthgrav}
In \textsc{SynthGrav}, a GW signal is constructed from a set of modes. Each mode is characterised by a time-dependent central frequency $f_c$.
The polarisation of each mode is determined individually and
the signal is constructed as a weighted sum of modes
\begin{align}
    h_{\times/+}(t) = \sum_i w_{\times/+}^i(t) A_{\times/+}^i(t),
\end{align}
where the $w_{\times/+}^i(t)$ indicate time-dependent weight
functions and the $ A_{\times/+}^i(t)$ denote the time-dependent amplitudes of the $i$-th mode.
When the final signal has been assembled as the sum of the individual modes,
it is possible to apply a global time-dependent weight--an envelope--to the total signal.

Each mode is constructed from a series of overlapping pulses, each consisting of coloured random noise. The length of each pulse is a key parameter and must be determined based on the problem. In general, the pulses should be much shorter than the total duration of the signal, but long enough to resolve the lowest frequency relevant for a given problem. For a pulse centred at $t^j$, we first define the central frequency $f^j_c = f_c(t^j)$. A sequence of random numbers is then generated from a uniform distribution, denoted as $ H_{\text{white}} $, with its Fourier Transform given by $\tilde{H}_{\text{white}}$. The subscript ``white'' highlights that these random numbers represent white noise. 
To colour the noise, we have to
choose the normalised power spectral
density (PSD) of the coloured noise.
First, we choose a function 
$S(f, f^j_c)$ centred around $f_c$. By default we use
a Gaussian for $S(f, f^j_c)$, but
the user can freely specify 
$S(f, f^j_c)$ and the software comes with a few predefined options (see the documentation accompanying the software).
The normalised PSD is then defined as 
\begin{align}
  S_{\text{norm}}(f,f_c) = S(f,f^j_c) / \sqrt{\langle S(f,f^j_c)^2 \rangle}, 
\end{align}
where $\langle S(f,f^j_c)^2 \rangle$ denotes the mean of $S(f,f^j_c)^2$.
The white noise is then weighted by the normalised PSD. This is done by multiplying the normalised PSD with the Fourier transform of the white noise $H_{\text{white}}$ in the frequency domain, which defines 
\begin{align}
    \tilde{H}_{\text{shaped}} = \tilde{H}_{\text{white}} \cdot S_{\text{norm}}(f).
\end{align}
Lastly, the shaped noise is converted back into the time domain by performing an inverse Fourier transform on $\tilde{H}_{\text{shaped}}$. We add the pulses together
to generate the mode in question. Depending on the choice of polarisation, the process
is repeated for the plus and cross components of $h$.

The user can freely determine the central frequency of each mode, but \textsc{SynthGrav}
comes with a set of supernova-specific modes based on the fitting formulas of \cite{Torres-Forne_19}. The fitting formulas allow the user to specify the key physical parameters for the time evolution and the software then calculates the time-dependent
central frequency of each requested mode. 

The code is publicly available at \url{https://github.com/haakoan/SynthGrav}. We
refer the reader to the accompanying documentation for a more in-depth explanation of the
code's functionality.

\section{An Optimal Scenario} \label{sec:optimal_scenario}
In the context of template-based searches,
the SNR for some filter $K(t)$ is given by
\begin{align} \label{eq:snrbasic}
    \SNR = \frac{\int\tilde{h}(f) \tilde{K}^{*}(f) \mathrm{d}f}
    {\Big(  \int \frac{1}{2}S_n(f) |\tilde{K}(f)|^2 \mathrm{d}f\Big)^{1/2}},
\end{align}
where $S_n(f)$ denotes the noise spectral density and $\tilde{K}^{*}(f)$ is the
complex conjugate of $\tilde{K}(f)$, which is the Fourier transform of $K(t)$. The Fourier transform of the signal $h(t)$ is represented by $\tilde{h}$. As is well-known in the GW astronomy literature, the
filter which maximises Eq.~\ref{eq:snrbasic} is the signal itself (see \cite{maggiore_07}
for a detailed derivation and discussion). 

The combination of semi-analytical methods and
numerical simulations has allowed for the creation of large template banks for compact
binary merger signals. The template banks used to detect compact binary mergers are almost
complete in large parts of the astrophysically motivated parameter space and contain templates that are almost
perfect matches to the observed signals. 
While computing the GWs emitted from binary mergers is a complicated procedure, the signals 
are relatively 
simple: they look like chirp signals that
grow louder the closer to merger the binaries get. The stochastic and noisy
GW signals expected from core-collapse supernovae are diametrically opposite to the signals from compact binary mergers.

It will never be possible to create complete template banks for core-collapse supernova signals, 
but the matched-filtering SNR assuming optimally-orientated detectors is commonly reported
in the literature because it provides a relatively easy way to gauge the detectability of
signal predictions from numerical simulations
(see, for example, the discussion in \cite{Andresen_17}).
Following this idea, we investigate how well filtering could perform under
the assumption of a perfect signal prediction and opportune detector orientations (relative to the source).
This analysis places an upper bound on the methodology presented in this paper. 

\begin{figure}
    \centering
    \includegraphics[width=1.0\linewidth]{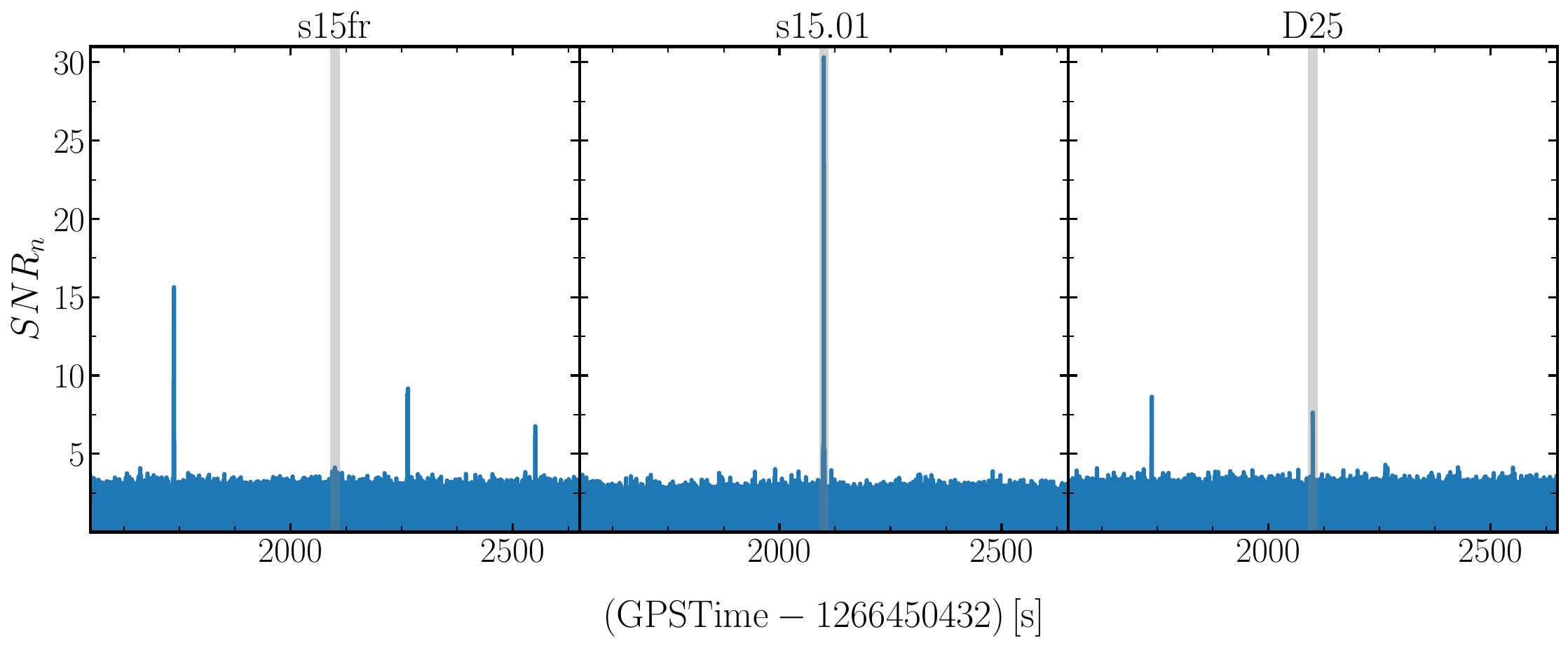}
    \caption{Network SNR for an event at 10 kpc using the signal itself as the filter. The left, middle, and right panels correspond to models s15fr, s15.01, and D25, respectively. The signal was injected 2100 s into the data frames starting at GPS time 1266450432. The shaded grey vertical region indicates the injection time. {The peaks in $\text{SNR}_n$ that do not align with the injection time are associated with non-Gaussian noise events (glitches).}}
    \label{fig:optimal}
\end{figure}

For one observer direction for each model, we select two 4096 s data frames from GWOSC (one for each detector) with a starting GPS time of 1266450432 and inject the corresponding signal into the data 2100 s after the start of the segments. We then perform template filtering using the injected signal itself as the template. The resulting $\SNR_n$ for each model is shown in Fig.~\ref{fig:optimal}: the left panel corresponds to model s15fr, the middle panel to model s15.01, and the right panel to model D25. In all cases, the signal was injected at the sky location (right ascension, declination) $= (24$ h, $35^\circ)$, with a polarisation angle of zero and assuming a source distance of 10 kpc. The shaded grey regions in Fig.~\ref{fig:optimal} highlight the injection time.
From Fig.~\ref{fig:optimal}, we see that the signal from model s15fr is indistinguishable from the background noise in $\SNR_n$, remaining below typical detection thresholds at this distance. In contrast, model D25 yields a more pronounced peak that is clearly visible above the noise. The largest $\SNR_n$ is obtained for model s15.01, which reaches a value of approximately 30.

Sampling 96 observer directions for the D25 and s15fr models, and the three available 
directions for the s15.01 model (observers situated along the $x$, $y$ and $z$ axes), we find
significant dependence of $\SNR_n$ on observer orientation. For D25, we obtain a mean $\SNR_n$ of 6.3, with values ranging from 3.8 to 7.9. For the s15fr model, the corresponding values are significantly lower, with a mean of 3.5 and a range between 2.7 and 4.6. In contrast, the s15.01 model yields substantially larger values, with a mean $\SNR_n$ of 28.7 and a range from 26.8 to 30.3. Using the optimal SNR as a detection statistic, these results indicate that a D25-like signal at 10 kpc would be detectable under favourable orientation, whereas the s15fr model would remain below a typical detection threshold at this distance. The signals from the s15.01 model yield the largest $\SNR_n$ values and would be readily detectable at 10 kpc. The comparatively large SNR obtained for s15.01 is likely attributable to the substantially longer duration of these waveforms, which increases the accumulated signal power. This suggests that accurately evaluating the detectability of GWs from core-collapse supernovae requires simulations that capture the full duration of the emission phase, and that template durations extending beyond those considered in the present template bank may improve detectability for long-lived signals.
We emphasise that these values represent optimal SNRs, as the injected signals are filtered with themselves. They therefore provide an upper bound on detectability for each model. The key question is how closely a practical template bank can approach this optimal scenario.

\section{Detecting and Reconstructing the Signal using a Template Bank}
\label{sec:detection_reconstruction}
\subsection{Detecting Events}
We now turn our attention to the realistic scenario in which we do not know
the exact supernova signal. Instead, we rely on an underlying theoretical understanding of core-collapse supernova GWs that we use to construct a template bank.
\begin{figure}
    \centering
    \includegraphics[width=1\linewidth]{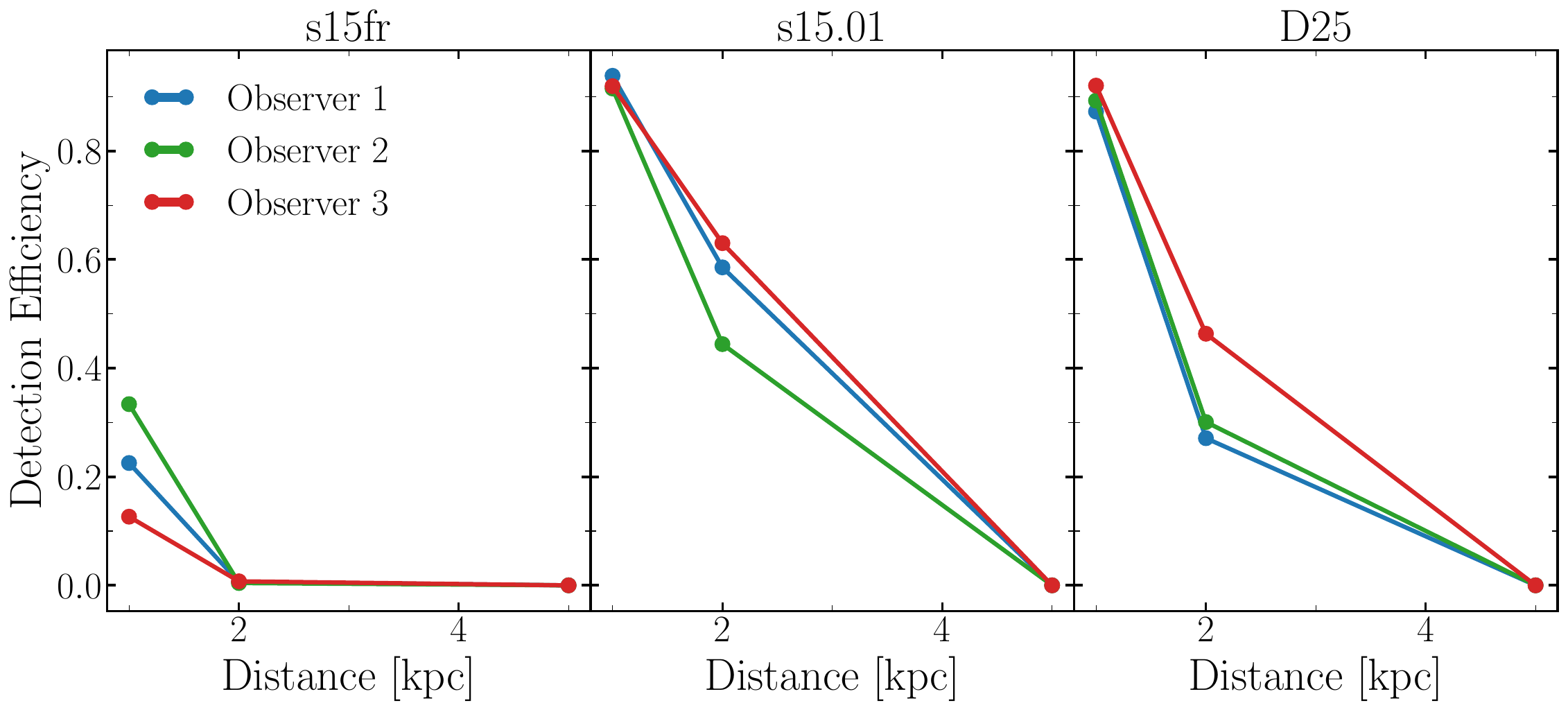}
    \caption{Detection efficiency as a function of source distance for models s15fr (left), s15.01 (middle), and D25 (right). For each model, results are shown for the three observer directions defined in the text. The dots indicate injections performed at distances of 1, 2, and 5 kpc. A network SNR threshold of 6 is used to define detection.}
    \label{fig:effec}
\end{figure}
The detection efficiency for the three models considered here is shown in Fig.~\ref{fig:effec}. The left, middle, and
right panels represent models s15fr, s15.01, and D25, respectively. For each model, results are presented for three observer directions. Observers 1 and 2 correspond to the same orientations for all models, located along the $z$- and $x$-axes, respectively, of the simulation coordinate system.
The third observer (Observer 3) was selected to probe a different viewing angle for each model. For model s15.01, only three directions were available. The third observer therefore corresponds to the $y$-axis of the simulation grid. For models s15fr and D25, the third observer was chosen at the directions $(\theta,\phi) = (\mathrm{\frac{\pi}{6},\frac{4\pi}{9}})$ and $(\theta,\phi) = (\mathrm{1.05,-1.57})$, respectively.
The detection efficiency is shown for a $\SNR_{n}$ threshold of 6.
Our choice of threshold can be
understood from Fig.~\ref{fig:optimal}; the correlation of the signal with pure detector noise leads to
an ambient SNR of $\sim 3\text{-}4$, so a threshold of 6 lies well above this background.
{Since we propose following up neutrino detections and do not perform a blind search, it is possible that a lower threshold could also be appropriate. However, a threshold of 6 provides a robust and conservative estimate.} 
Based on Fig.~\ref{fig:optimal}, we expect to pick up several false triggers with a threshold of 6. {The noise spikes seen in Fig.~\ref{fig:optimal} that do not coincide with the injection time are due to noise glitches.}

However, as mentioned above, we assume that false triggers can be eliminated since they will not coincide with the arrival time of the neutrinos.
{We disregard a total of 933 triggers that do not align with the injection time, corresponding to a \reply{false-alarm rate (FAR) of $\sim185$ per day} (see Section~\ref{sec:false_alarms} for more details regarding false alarms).}
It is, of course, possible that a glitch occurs at exactly the same time as a real event, but we do not consider this scenario in our current analysis.

From Fig.~\ref{fig:effec}, we see that all of the signals are effectively undetectable
at a distance of 5 kpc with the templates used in this study.
For the s15fr model, the detection efficiency is relatively low even at 1 kpc and shows a strong dependence on observer direction. At 1 kpc, the efficiency ranges from 0.13 for the $x$-axis observer to 0.33 for the $z$-axis observer, with the third observer direction yielding an intermediate value of 0.23. At 2 kpc, the efficiency drops significantly for all orientations, falling to below 1\%, which means that the signal from this model is effectively undetectable beyond distances of order $\sim2$ kpc with our approach.
In contrast, the s15.01 model shows substantially higher detection efficiencies and a weaker dependence on orientation. At 1 kpc, efficiencies lie between 0.87 and 0.92, with the $x$-axis observer yielding the highest value. At 2 kpc, the efficiency decreases but remains significant, ranging from 0.27 to 0.46, again with the $x$-axis observer providing the most favourable orientation.
The D25 model exhibits behaviour broadly similar to s15.01. At 1 kpc, the detection efficiency is very high for all observers, ranging between 0.91 and 0.94, with the highest value obtained for observer 3. At 2 kpc, the efficiency decreases, but the signal remains detectable, with detection efficiencies between 0.44 and 0.63 and a relatively small variation between observer directions.

Comparing the three models, s15fr shows both the lowest detection efficiency and the largest variability with observer orientation. At 1 kpc the difference between the best and worst observer directions is approximately 0.20, whereas for s15.01 and D25 the corresponding spreads are only 0.05 and 0.03, respectively. This means that the detectability of the s15fr signal is both weaker overall and more sensitive to viewing angle than the other two models. The strong emission around 100 Hz seen in model s15fr is associated with the SASI and is known to show a stronger dependence on the observer direction than the higher-frequency emission \cite{Andresen_17}. At small distances, some of the templates can capture part of the SASI signal, which together with the directional dependence of the SASI signal likely explains the variation in detection efficiency as a function of observer direction.

The fact that model s15fr is difficult to detect using our template bank
shows that the approach works largely as expected. As can be seen in
Fig.~\ref{fig:signal} and discussed above, the signal from s15fr is
relatively weak, with the exception of the emission generated by the SASI. 
The high-frequency component that our template bank is designed to capture is
weak and subdominant in this model. Consequently, the low detection
efficiencies obtained for s15fr are not surprising.
At the same time, this highlights an important challenge associated with
supernova signals. Some models predict signals that differ significantly
from the standard picture, and there remains considerable variation within
the signal class. Constructing a complete template bank is
therefore challenging and will rely on continued progress in numerical
simulations and theoretical predictions of the GW signals.
It should also be noted that model s15fr is not necessarily representative
of a typical supernova model. In this simulation, rotation aided both the
growth of the SASI and the explosion itself, while at the same time
suppressing the high-frequency emission \cite{Andresen_19}.

The detection efficiencies obtained with our template bank follow this same general
trend obtained for the optimal scenario discussed above: model s15fr is difficult to detect, whereas models D25 and s15.01 yield significantly higher detection efficiencies.
However, the detection efficiencies obtained for s15.01 are noticeably
lower than the optimal SNR values would predict. One possible
explanation is the substantially longer duration of the s15.01
waveforms. Since the SNR accumulates signal power over
time, the comparatively short templates used in
our study will miss part of the signal for s15.01. This suggests
that extending the template duration could improve the performance of
the template bank for long-lived supernova signals.

{From the definition of the fitting factor, it follows that
$\text{SNR}_\mathrm{bank} \approx \mathrm{FF} \times \text{SNR}_\mathrm{opt}$, where
$\text{SNR}_\mathrm{opt}$ is the matched-filter SNR achievable with the signal itself
as the template and $\text{SNR}_\mathrm{bank}$ is the SNR achieved with the
best-matching template in the bank \citep{Apostolatos_95,Owen_99}.
Scaling the optimal SNRs of Section~\ref{sec:optimal_scenario} to the 
distances used in this section, we estimate expected SNR ranges for our 
template bank at $1$~kpc and $2$~kpc.
At $1$~kpc we find $\text{SNR}_\mathrm{bank} \approx 14$--$36$ 
for D25, $\text{SNR}_\mathrm{bank} \approx 80$--$124$ for s15.01, and 
$\text{SNR}_\mathrm{bank} \approx 2$--$9$ for s15fr.
At $2$~kpc the expected SNRs become 
$\text{SNR}_\mathrm{bank} \approx 7$--$18$ for D25, 
$\text{SNR}_\mathrm{bank} \approx 40$--$62$ for s15.01, and 
$\text{SNR}_\mathrm{bank} \approx 1$--$5$ for s15fr. The estimates for D25 and 
s15.01 lie above our detection threshold of $\mathrm{SNR}_n = 6$ at both 
distances, consistent with their high empirical detection efficiencies of 
$\sim 90\%$ at $1$~kpc. The estimated SNR range for s15fr includes our 
threshold of 6, consistent with its substantially lower detection 
efficiency of $\sim 10$--$30\%$. Based on the estimated SNR values, we 
would expect higher detection fractions at $2$~kpc for D25 and s15.01 
than what we obtain. However, these estimates do not account for variations in 
the noise floor or for the antenna patterns of the detectors.}

It is instructive to place our results in the context of existing
searches for GWs from core-collapse supernovae.
However, direct comparisons between methods should be treated with
caution. Our study considers a follow-up search scenario and does not
impose a fixed FAR, whereas most existing studies
evaluate blind searches and report detection efficiencies at a
specified FAR. Differences in the detector noise used, the injected
waveforms, and the treatment of glitches further complicate direct
comparisons.
With these caveats in mind, our detection efficiencies are broadly
consistent with the general conclusion of model-independent burst
searches that Galactic core-collapse supernova signals may be detectable in favourable
cases. For example, Szczepańczyk et al.~\cite{Szczepanczyk_21} study
core-collapse supernova detectability with coherent WaveBurst using
detector noise scaled to expected O5 sensitivities
\cite{LIGO_T2000012} and find that neutrino-driven explosions can be
detectable throughout a significant fraction of the Milky Way.
An optically targeted search during O3~\cite{Szczepanczyk_24}
reports that waveforms similar to those used in this work can be
detected with an efficiency of $\sim50\%$ at 2~kpc (see Fig.~6 of
\cite{Szczepanczyk_24}).
A meaningful comparison between excess energy methods and
template-based approaches would require injecting the same waveforms,
using identical detector noise, and evaluating the detection
efficiency at a fixed FAR. Such a study is beyond the scope of the
present work. Nevertheless, our results suggest that template-based
methods may provide a useful complementary approach in follow-up
searches triggered by neutrino or electromagnetic observations of a
Galactic supernova.

\subsection{Reconstructing the Signal} \label{sec:recon}
\begin{figure}
    \centering
    \includegraphics[width=1\linewidth]{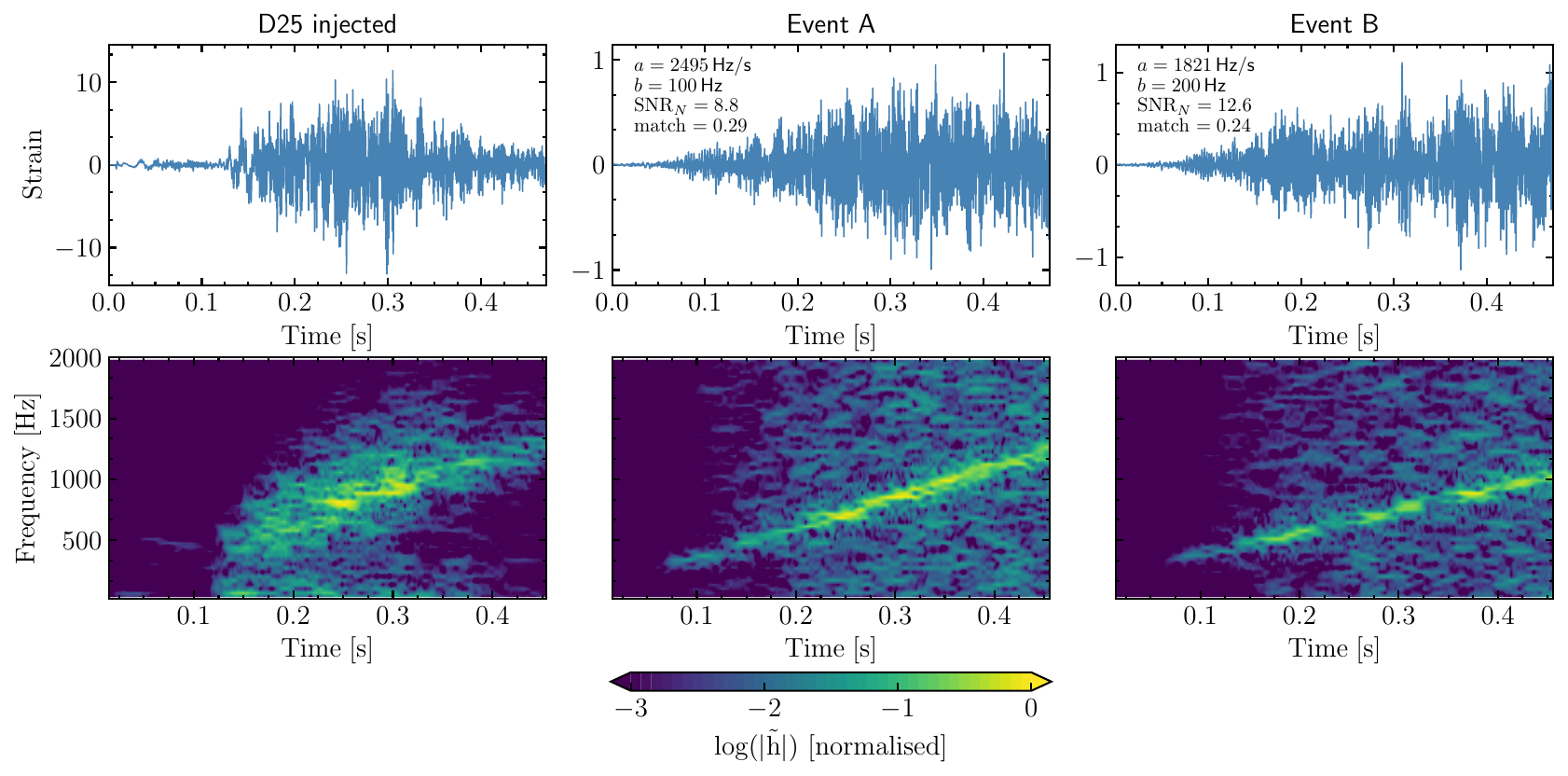} \\
    \includegraphics[width=1\linewidth]{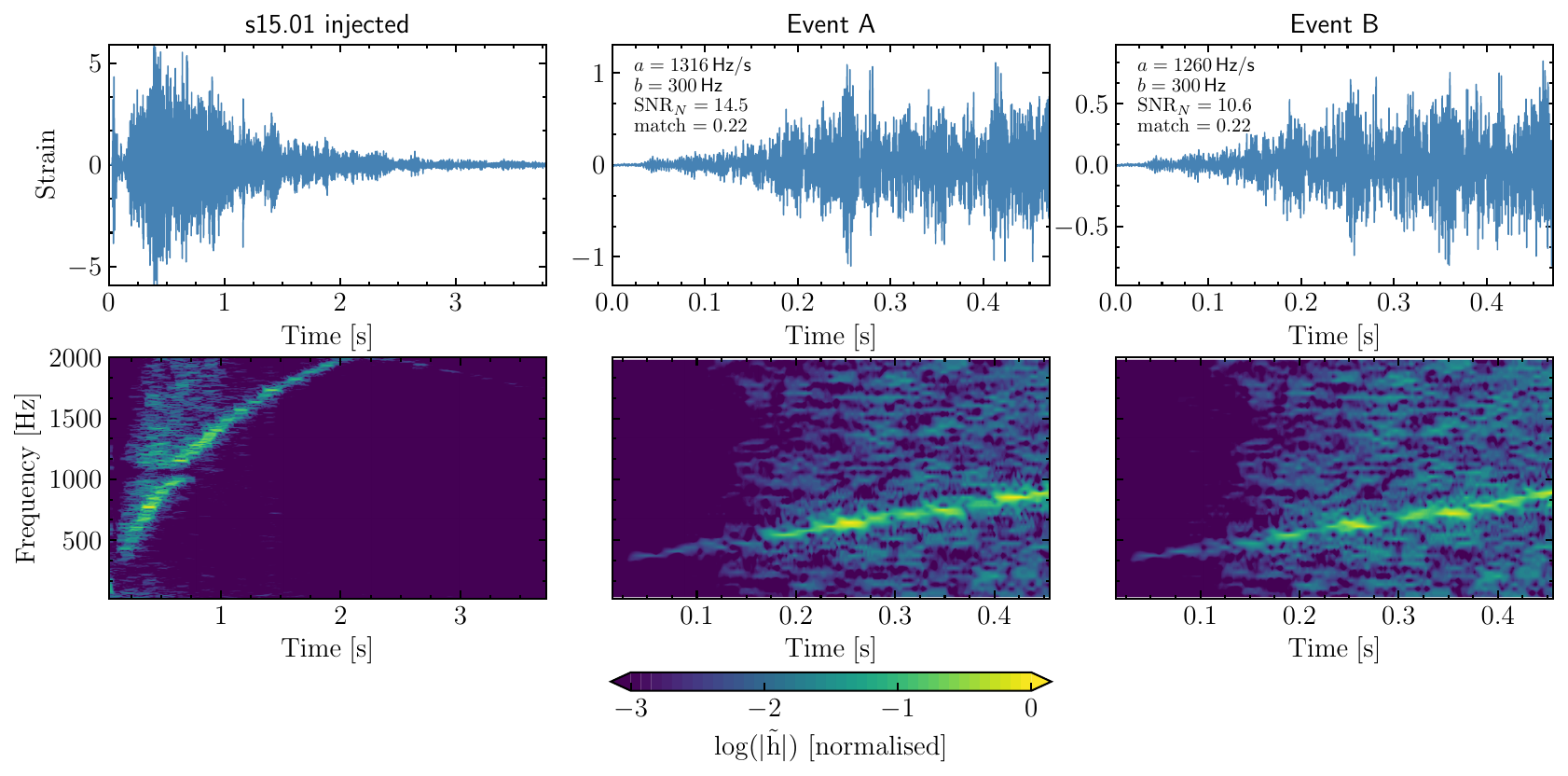} \\
    \includegraphics[width=1\linewidth]{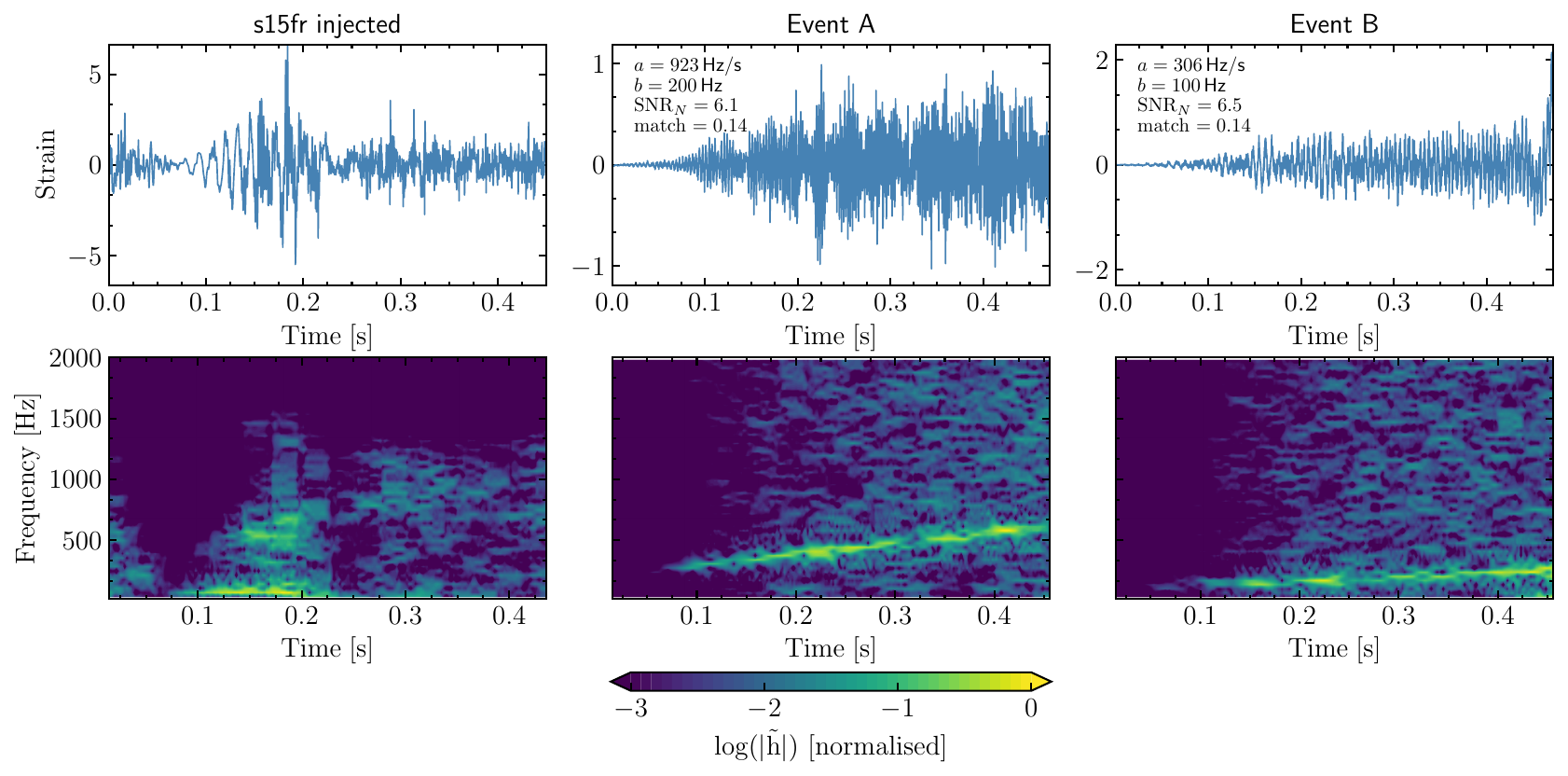}
    \caption{Injected signals at $1$~kpc and the recovered templates for two
    arbitrarily selected events from the search, for each of
    the three models considered in this work: D25 (top block), s15.01
    (middle block), and s15fr (bottom block). For each model, the upper row
    of panels shows time series and the lower row shows the corresponding
    spectrograms. Within each block, the columns show, from left to right,
    the injected signal, the recovered template for event A, and the
    recovered template for event B. Each recovered-template panel is
    annotated with the template parameters $(a, b)$, the network
    signal-to-noise ratio $\mathrm{SNR}_n$ returned by the search, and the
    match $m(s, h_\mathrm{rec})$ between the injected signal and the
    recovered template, computed via Eq.~\ref{eq:match}. Spectrograms are
    normalised by the peak value in each panel and shown on a logarithmic
    scale.}
\label{fig:reconstructed}
\end{figure}

The reconstructed signal is defined as the template that yields the
highest $\SNR_n$. 
{Fig.~\ref{fig:reconstructed} shows the recovered signals for two arbitrary events
for the three models considered in this work. 
The recovered $\SNR_n$ values lie above our detection threshold of $\SNR_n = 6$ in all cases.
The match values $m(s, h_\mathrm{rec})$ between the injected signal and the recovered
templates lie between $0.14$ and $0.29$. Note that the match is computed for a fixed observer direction and does
not include the antenna response at the injection's sky position, while
$\SNR_n$ does. Therefore, a recovery with a higher $\SNR_n$ can correspond to
a slightly lower match (or vice versa).

For the s15.01 model, the recovered templates for the two events have
nearly identical parameters and capture the main signal component visible in the
signal spectrogram in the first second of the waveform. For D25, the
two recovered templates differ in their precise parameter values but
both lie in the same region of the bank and pick up the rising-frequency
feature visible in the D25 signal between $\sim 0.1$ and $\sim 0.4$~s.
The templates that match the signal from model s15fr have small $a$ and $b$
values, which indicate that the template picks up part of the SASI signal and potentially
some of the emission at higher frequencies.}

\begin{figure*}
    \centering
    \includegraphics[width=1\linewidth]{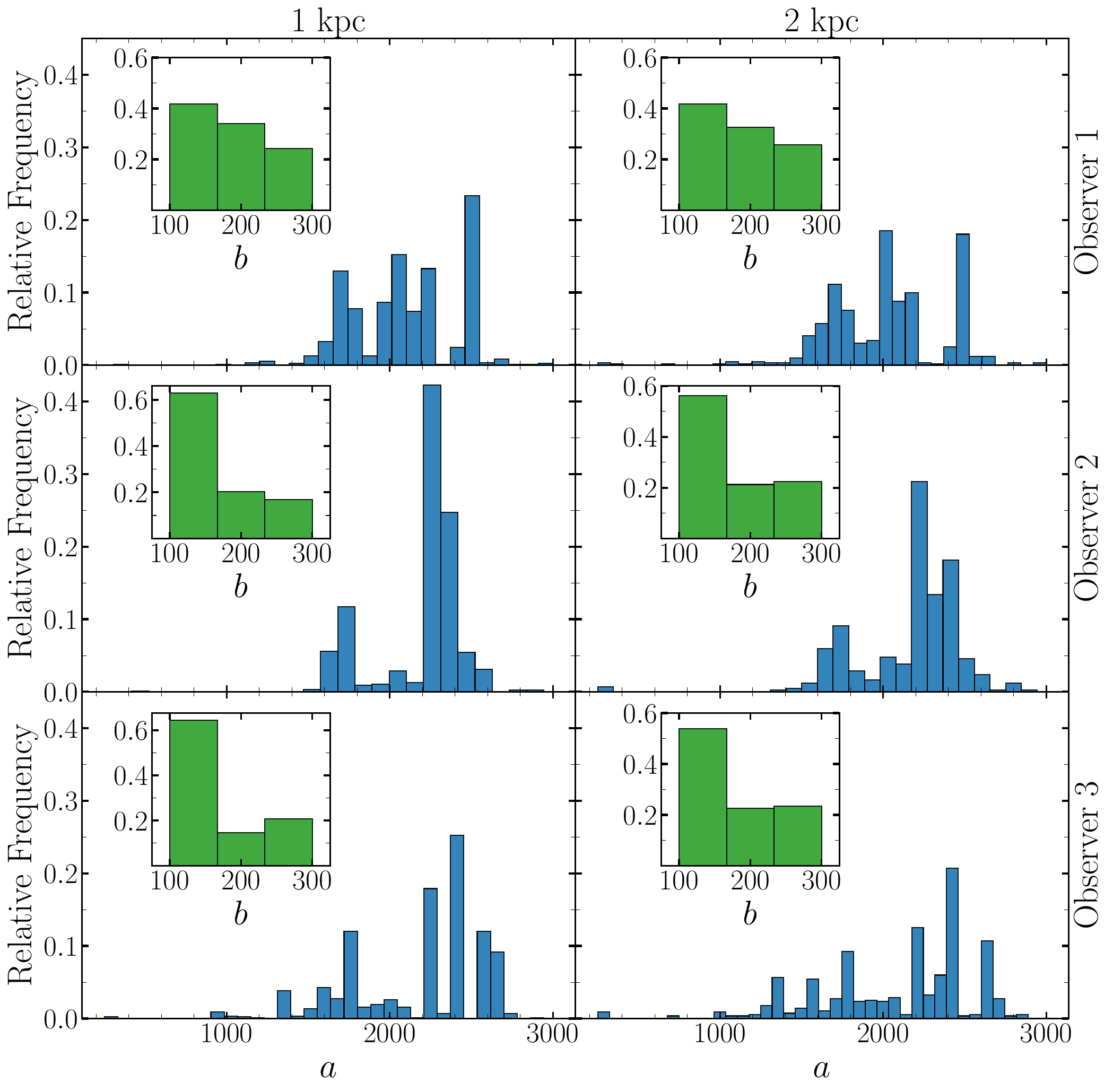}
    \caption{Distribution of reconstructed parameters ($a$ and $b$) for the best-matching templates for model D25. The three rows correspond to the three observer directions considered in this study. The left and right columns show results for signals injected at distances of 1 kpc and 2 kpc, respectively. Blue bars represent reconstructed $a$-values, while green bars in the inset show reconstructed $b$-values. The parameter $b$ is given in Hz and $a$ in Hz {$\cdot$} s$^{-1}$.}
    \label{fig:rec_d25}
\end{figure*}

\begin{figure*}
    \centering
    \includegraphics[width=1\linewidth]{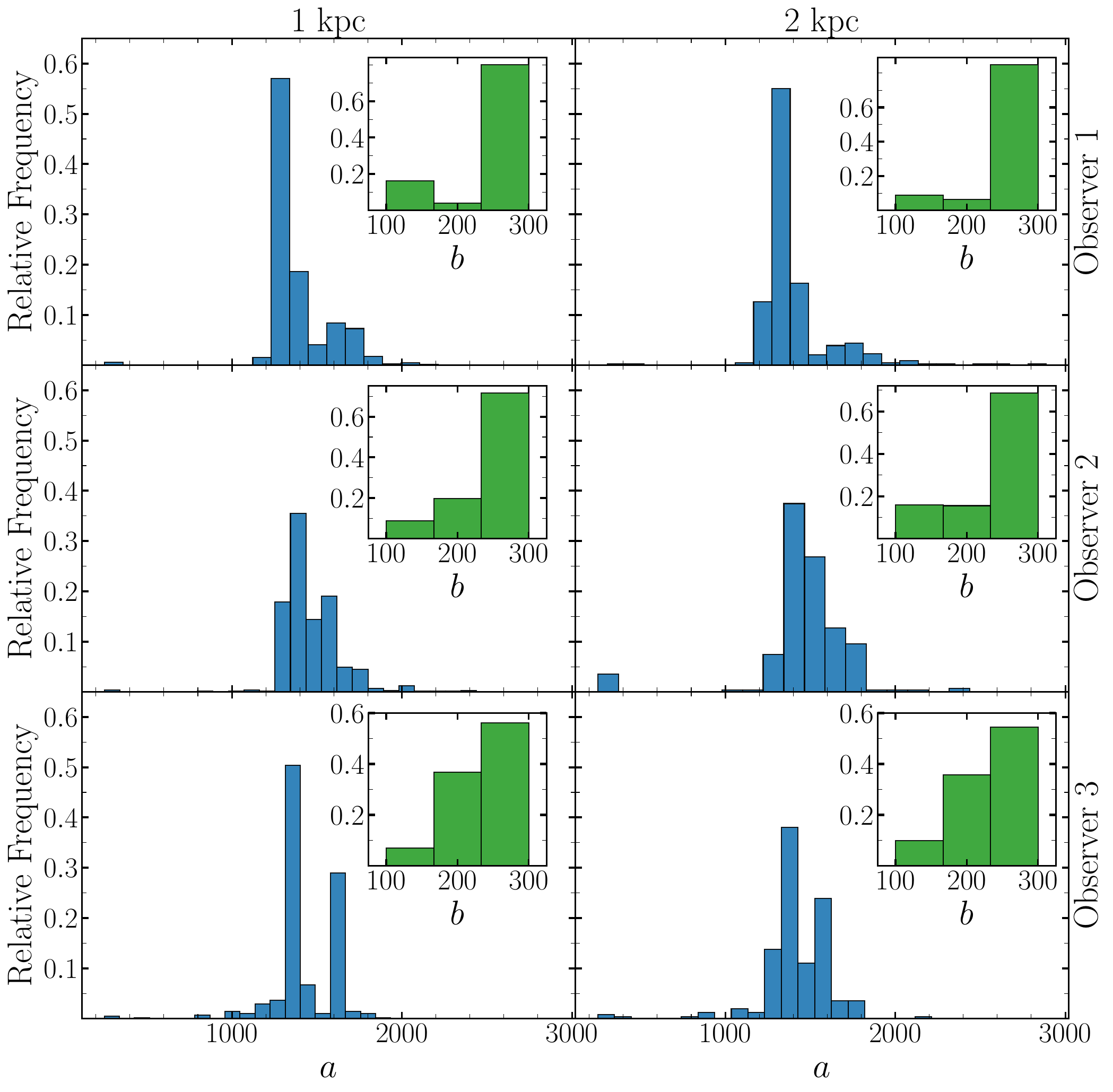}
    \caption{Distribution of reconstructed parameters ($a$ and $b$) for the best-matching templates for model s15.01. The three rows correspond to the three observer directions considered in this study. The left and right columns show results for signals injected at distances of 1 kpc and 2 kpc, respectively. Blue bars represent reconstructed $a$-values, while green bars in the inset show reconstructed $b$-values. The parameter $b$ is given in Hz and $a$ in Hz $\cdot$ s$^{-1}$.}
    \label{fig:rec_s15}
\end{figure*}

\begin{figure*}
    \centering
    \includegraphics[width=1\linewidth]{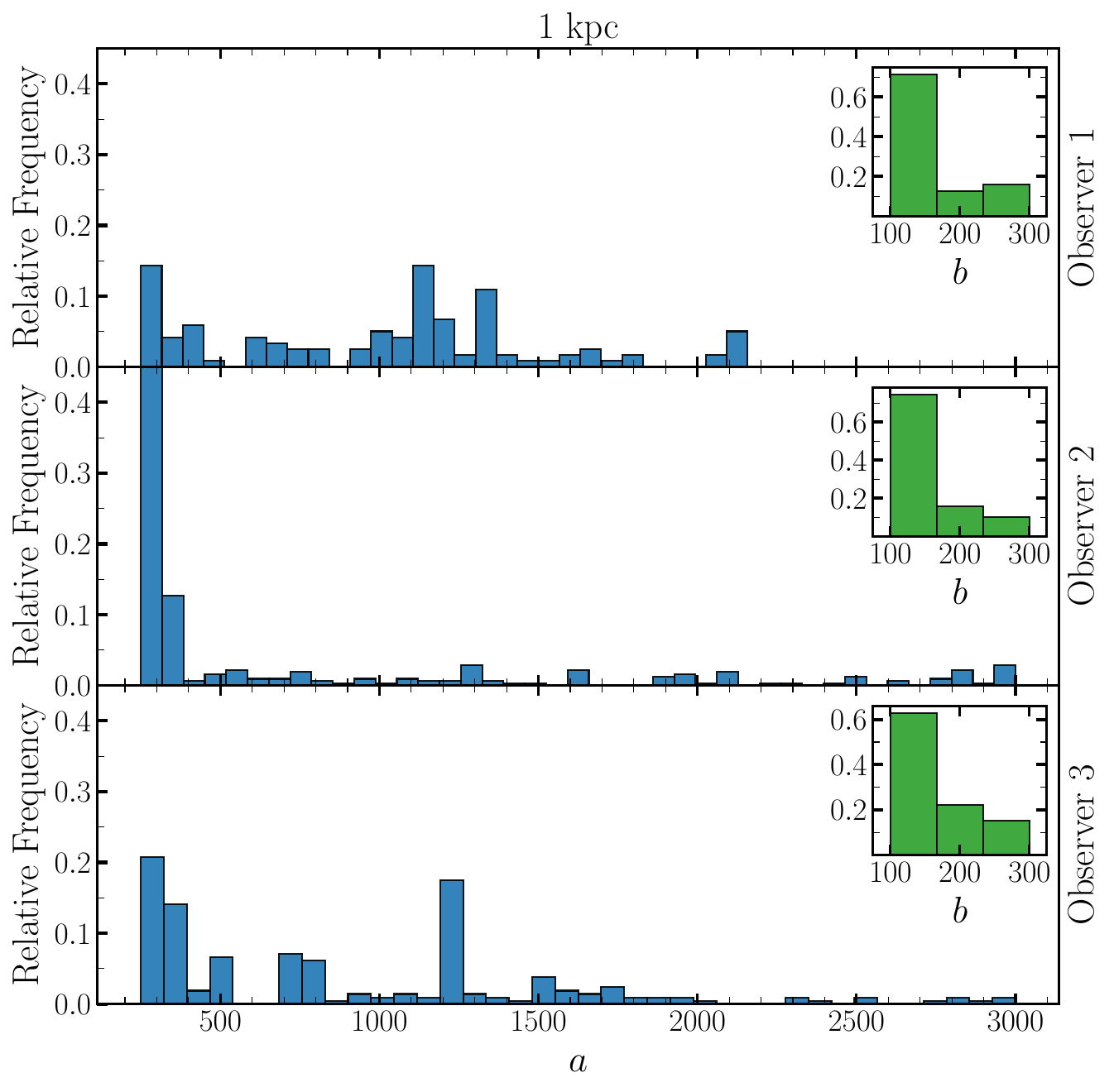}
    \caption{Distribution of reconstructed parameters ($a$ and $b$) for the best-matching templates for model s15fr. The three rows correspond to the three observer directions considered in this study. The figure shows results for signals injected at distances of 1 kpc. Blue bars represent reconstructed $a$-values, while green bars in the inset show reconstructed $b$-values. The parameter $b$ is given in Hz and $a$ in Hz $\cdot$ s$^{-1}$.}
    \label{fig:rec_s15fr}
\end{figure*}

{It is clear from Fig.~\ref{fig:reconstructed} that the pipeline recovers a range of values for
the underlying template parameters.  }
We show the distribution of reconstructed $a$ and $b$
values in
Figs.~\ref{fig:rec_d25}, \ref{fig:rec_s15}, and \ref{fig:rec_s15fr}. In each figure, the rows
correspond to the three observer directions considered in the analysis.
For the s15.01 and D25 models, the left column shows results for an
injection distance of 1 kpc, while the right column corresponds to
2 kpc. For the s15fr model, only a single column is shown since our pipeline essentially failed to detect the signal from model s15fr at
2 kpc. In each panel, the blue bars in the main plot represent the reconstructed
$a$ values, while the green bars in the inset show the reconstructed $b$
values. Here, $a$ and $b$ determine the time evolution of the central frequency of the template, 
$f_c(t) = at + b$ (see Section~\ref{sec:templates}).
For both the main panels and the insets, the $y$-axis gives the
relative frequency of a reconstructed value, defined as the number of
events reconstructed with a given parameter divided by the total number
of recovered events.

The reconstructed parameter distributions for model D25 are shown in
Fig.~\ref{fig:rec_d25}. Overall, the reconstructed $a$ values are centred
around $a \sim 2000$--$2700$ for all three observer directions.
The reconstructed $a$ distributions are relatively narrow, particularly
for Observer~2, where most recovered events concentrate around
$a \sim 2200$--$2400$. Observers 1 and 3 show slightly broader
distributions, but still peak in the same range. The reconstruction for
Observer~2 shows hints of a bimodal distribution, while the other two
observer directions show smaller secondary peaks. Increasing the
injection distance from 1 to 2 kpc broadens the distribution.
For Observers~2 and~3, the majority of reconstructed events favour
$b = 100$. For Observer~1, $b = 100$ is also the most commonly
reconstructed value, although the distribution is relatively flat
compared to the other two observers. The reconstructed $b$
distributions show little variation with source distance.

The reconstructed parameter distributions for model s15.01 are shown in
Fig.~\ref{fig:rec_s15}. Overall, the reconstructed $a$ values are centred
around $a \sim 1400$--$1700$ for all three observer directions.
At an injection distance of 1 kpc, the reconstructed $a$ distributions
are relatively narrow and show clear peaks within this range.
Observer~1 shows the most concentrated distribution, while Observer~2
shows a slightly broader spread but still peaks at similar
values and Observer 3 shows a secondary peak at $\sim1600$. 
Increasing the injection distance from 1 to 2 kpc broadens the
distribution for all observers, although the reconstructed values
remain centred roughly around a large peak at $\sim 1400$.
For all observer directions, $b = 300$ is the most frequently recovered value at both
distances. Observers~1 and~2 show particularly strong peaks at this
value, while Observer~3 exhibits a somewhat broader distribution with
a larger fraction of events reconstructed with $b = 200$. As for the
$a$ parameter, the reconstructed $b$ distributions show only minor
changes when the injection distance is increased from 1 to 2 kpc.

We show the reconstructed parameter distributions for model s15fr in
Fig.~\ref{fig:rec_s15fr}.
Note that we only show the results for a distance of 1 kpc, 
since the signal is not efficiently recovered at larger
distances. For Observers~1 and~3, the reconstructed $a$ values span a
wide range and do not show a clear dominant peak. The distributions are
broad and contain several smaller clusters. A significant fraction of
events are reconstructed with relatively small $a$ values around
$a \sim 300$--$400$, for Observers~2 and~3. For
Observer~2 we find a pronounced peak at $a \sim 300$.
For all three observer directions, the reconstructed distributions
strongly favour $b = 100$, which accounts for the majority of recovered
events.

From Fig.~\ref{fig:signal}, we estimate the slope of the high-frequency
component of the signal to be approximately $2700$ Hz/s for model D25,
while the corresponding slopes are $\sim1500$ Hz/s for s15.01 and
$\sim2000$ Hz/s for s15fr. The reconstructed $a$ values obtained with
our template bank are broadly consistent with these estimates for the
D25 and s15.01 models. Going by the largest peak, the error
in the recovered $a$-value is on the order of $\sim 100$, which
is an error of around $10-20$\%.
For the s15fr model,
the slope is both difficult to estimate from the signal and more 
difficult to constrain since the high-frequency
component is comparatively weak and the signal is largely dominated by
the nearly stationary SASI emission around $\sim100$~Hz.

A number of recent studies have explored the recovery of the
time-frequency evolution of core-collapse signals using excess energy
pipelines. For example, \cite{Casallas-Lagos_23} used machine learning
techniques to reconstruct the time-frequency structure of injected
signals recovered with the excess energy pipeline
\textsc{Coherent WaveBurst} \citep{klimenko_16,klimenko_21}. 
The accuracy reported by \cite{Casallas-Lagos_23} is comparable to the spread of 
reconstructed a values we obtain here, although their results depend on the specific waveform considered.
Similarly,
\cite{Powell_22} show that for signals with sufficiently large network
SNR, the frequency evolution of injected bursts can be recovered to
within $\sim100$~Hz (see their Fig.~5).

Several factors can influence which template is identified as the reconstructed signal, including detector noise, partial overlap between templates in time-frequency space, and the clustering procedure used to identify the maximum $\SNR_n$ within each time window. 
Improving the accuracy of the reconstruction would likely require
a larger and more densely sampled template bank, as well as
dedicated parameter-estimation methods capable of exploring the
template space more systematically.

Reconstructing the central value of the main signal component
and its slope is particularly interesting for core-collapse
signals because it is set by the properties of the
proto-neutron star (PNS). It has been shown that
the central frequency of the emission can be expressed
as
\begin{align} \label{eq:cf}
f_{\text{GW}} &= \frac{1}{2\pi} \frac{GM_{\text{PNS}}}{R_{\text{PNS}}^2}
\sqrt{1.1 \frac{m_n}{\langle \epsilon_{\bar{\nu}} \rangle}} 
\left[1 - \frac{GM_{\text{PNS}}}{R_{\text{PNS}}c^2}\right]^2,
\end{align}
where $M_{\text{PNS}}$ is the mass of the PNS,
$R_{\text{PNS}}$ is the radius of the PNS, $\epsilon_{\bar{\nu}}$
is the average anti-electron neutrino energy \citep{muller_13}. The constants
$m_n$, $c$, and $G$ are the neutron rest mass,
the speed of light, and Newton's constant, respectively. 
The average energy of anti-electron neutrinos can be measured
from neutrino observations, which leaves two unknowns: the mass and radius of the PNS. Together, the PNS mass and PNS radius determine the compactness of the PNS, which is a measure of the properties of the underlying equation of state.

\section{False Alarms} \label{sec:false_alarms}
In this work, we focus on a follow-up search scenario and we
have, therefore, not considered false alarms so far. Nevertheless, 
it is useful to briefly discuss
false triggers in the context of our analysis.

We define a false alarm as an event that is not associated with an
injected signal but produces a network SNR above our chosen threshold
of $\SNR_n = 6$. The FAR is defined as the number of such events per
unit time. Since the overlap between detector noise and individual
templates varies across the template bank, the FAR depends on the
number of templates included in the search. In contrast to the case
of stationary Gaussian noise, where the FAR can be predicted from the
statistical properties of the SNR, real detector data
is neither Gaussian nor stationary. Consequently, the FAR must be
measured empirically (see, e.g., \cite{nitz_17,kumar_24}).

\reply{We estimate the FAR by time-shifting the L1 SNR time series relative
to H1 by 100 distinct shifts ranging from 10 to 307 seconds. The shifts
are integer multiples of 1~s, matching the 1~Hz sampling of the SNR time
series used in our analysis. 
These shifts are well above the H1--L1 light-travel time of
approximately 10~ms, which ensures that the background is free from
any astrophysical signal. This procedure yields approximately one year
of effective background.}

\begin{figure}
    \centering
    \includegraphics[width=1\linewidth]{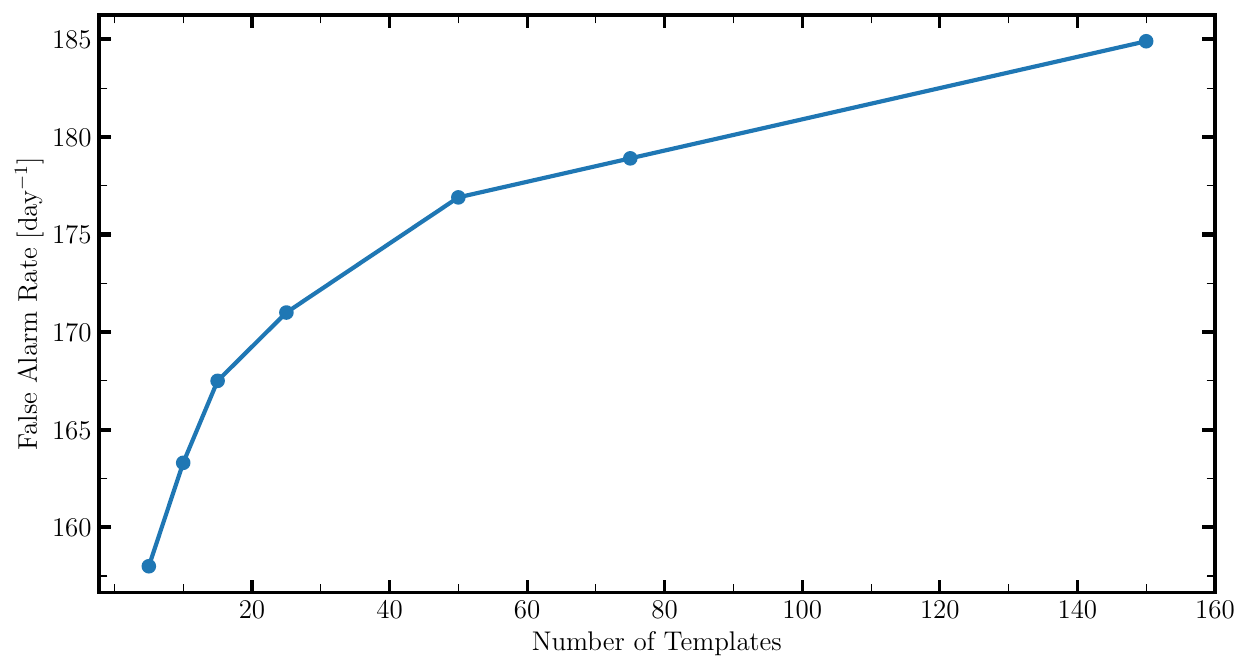}
    \caption{FAR as a function of the number of templates in the template bank. The blue dots represent actual data points. }
    \label{fig:falarmvsntemp}
\end{figure}

Fig.~\ref{fig:falarmvsntemp} shows the FAR as a
function of the number of templates included in the analysis. 
\reply{Using
the full template bank consisting of 150 templates, we obtain a FAR
of $184.9$ events per day. The FAR increases with the number of
templates, although the dependence is relatively weak: the difference
between using five and 150 templates is approximately $15\%$, while
the difference between 50 and 150 templates is roughly $4\%$.}
\begin{figure*}
    \centering
    \includegraphics[width=1\linewidth]{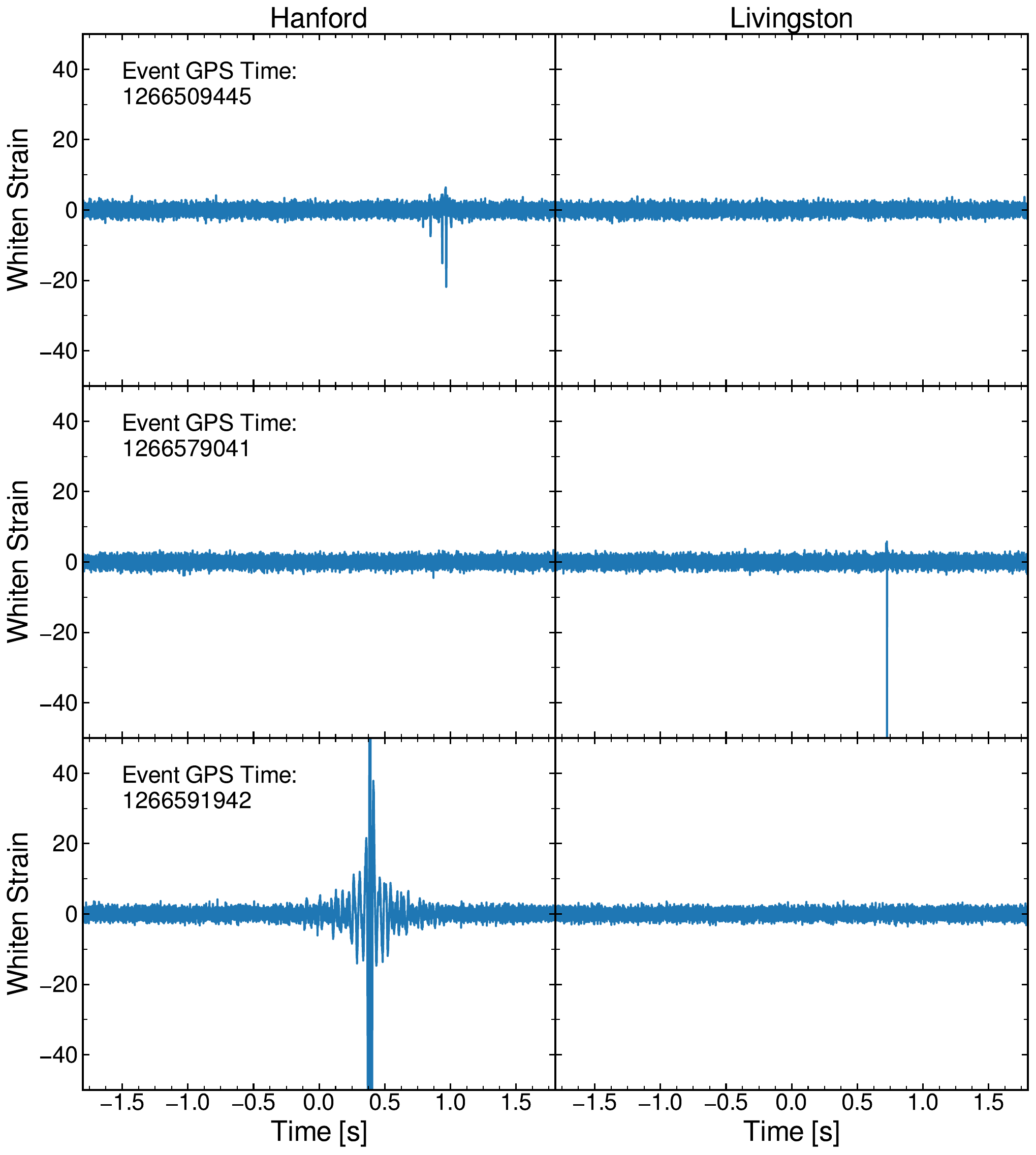}
    \caption{Examples of noise glitches that generated false alarms during the analysis. Each row displays a different noise glitch resulting in a false alarm. The GPS time for each event is indicated in the upper left corner of the left panel.
    The left column represents
    the data stream from Hanford. Data from Livingston are shown in the right column. The time axis for each panel is a 3 s wide window around the noise trigger.
    The data shown has been whitened using \textsc{gwpy.timeseries.whiten()}.}
    \label{fig:ftriggers}
\end{figure*}

{The FAR reported above is relatively high, which means
that our approach, in its current form, cannot be used to confidently
detect GWs from core-collapse supernovae without a coincident neutrino
detection. Separating real events from noise triggers would require a
substantially higher $\SNR_n$ threshold, which would in turn reduce
the detection efficiency. Moreover, raising the threshold alone would
not eliminate the loudest noise triggers: some of the largest $\SNR_n$
values produced by our pipeline are caused by detector glitches, which
would survive any threshold tuning.  The method is therefore, in its current form,
not suited to be run as a blind search.}

Most false triggers occur within $\sim1$~s of clearly identifiable
noise glitches in the detector data. Examples of such glitches are
shown in Fig.~\ref{fig:ftriggers}. A smaller number of false triggers
cannot be directly associated with obvious glitches. In this work, we
did not attempt to veto glitches or perform a detailed noise study
targeted at supernova signals. Existing glitch-removal techniques
\citep{Bondarescu_23,Tolley_23} could likely reduce the false trigger
rate significantly.

For galactic supernovae, gravitational-wave searches will likely be
performed as part of a multi-messenger analysis in which neutrino or
electromagnetic observations provide an external trigger time
\citep{Azfar_24}. In such a scenario, the relevant background is the
probability of a noise trigger occurring within the time window of
the external trigger rather than the FAR of a blind search. A detailed
evaluation of the background for such triggered searches is beyond
the scope of the present work.

\section{Conclusions} \label{sec:conclusions}
In this work, we explored the feasibility of detecting GWs
from core-collapse supernovae with template-based methods
via signal injections into data from the LVK O3b run.
Using the newly developed template generation code
\textsc{SynthGrav}, we constructed a template bank of synthetic
core-collapse supernova GW signals. \textsc{SynthGrav} is specifically
designed to reproduce the time–frequency characteristics observed
in core-collapse supernova signals. Crucially, while the synthetic
signals do not match the injected signals perfectly, the overall
morphology of the templates resembles the injected signals well
enough to enable the use of filtering methods.

We studied the detectability of three signals from numerical
core-collapse simulations (models D25, s15.01, and s15fr).
Under optimal conditions, using the signal itself as the template,
we find that detectability depends strongly on the signal.
The s15.01 model yields the largest network SNR values and would be
readily detectable at a distance of 10 kpc. The D25 model can reach
network SNR values above typical detection thresholds under favourable
observer orientations, while the model s15fr would be difficult to detect
at a distance of 10 kpc.

Using a template bank consisting of 150 templates constructed with
\textsc{SynthGrav}, we find detection efficiencies of
$\sim90\,$\% at a distance of 1 kpc for the models D25 and s15.01,
decreasing to $\sim 30$–$60$\% at 2 kpc depending on the observer
direction. For model s15fr the detection efficiency is
significantly lower, reaching only $\sim 10$–$30$\% at 1 kpc and
dropping to below 1\% at 2 kpc. 
{The signal from model s15fr is not detected because our template bank does
not capture the signature of emission driven by strong SASI activity. Including this model
in our analysis highlights an important limitation of template-based searches for
core-collapse supernova signals. On the other hand, the majority of numerical
models predict the strongest emission in the signal component we study here, at least in the
absence of rapid rotation.}
In all cases, no signals were detected
at a distance of 5 kpc. We find $\sim5-20$\% variations in the
detectability as a function of observer orientation.

For the models s15.01 and D25, where the high-frequency component of the signal is
prominent, the reconstructed template parameters broadly reproduce
the time–frequency evolution of the injected signals. However,
significant scatter remains in the reconstructed parameters.
This indicates that the template-based approach
can provide qualitative information about the signal properties and, therefore, the
underlying physical properties of the supernova core. However, 
precise parameter reconstruction might require a larger template bank than
what we consider here, as well as more robust parameter estimation methods.

{The false alarm rate of our analysis with the full bank is
$\sim 180$ events per day, dominated by detector glitches. In its
current form, our approach is therefore best suited to triggered
follow-up searches in which neutrino observations
supply the time of the event, rather than blind all-sky searches.
Increasing the detection threshold can eliminate some false triggers, but
it would in turn reduce our detection efficiency. At the same time, several loud noise
events associated with glitches would survive regardless of what threshold we set.
Reducing the false alarm rate sufficiently to enable a blind search
would require glitch vetoes.}

Looking to the future, several improvements could enhance the accuracy
and sensitivity of the method proposed in this paper.
\begin{enumerate}
\item \textit{Improving the template bank.} Developing a more
comprehensive template bank that includes a broader range of possible
supernova signals would improve detection prospects, particularly for
progenitors that exhibit additional signal components such as emission
associated with SASI activity. Furthermore, the simple linear
central-frequency evolution used in this work should be replaced with
more realistic functional forms. Torres-Forn\'e et al.
\cite{Torres-Forne_19} derived polynomial expressions that relate the
central frequencies of various modes to the underlying physical
properties of the supernova core. Their prescription is implemented in
\textsc{SynthGrav} and can be used in the future to generate more
sophisticated templates. {It will also be important to develop
more complete template banks, with higher fitting factors than we
obtained for our template bank.}

\item \textit{Refining the template filtering methodology.}
Improvements to the template-based filtering procedure, such as better
PSD estimation in shorter time segments and improved clustering of SNR
peaks, could further increase the efficiency of the method.

\item \textit{Exploring a broader set of supernova signals.} In this
work, we have focused primarily on the high-frequency component of the
GW signal, often attributed to oscillation modes of the PNS. 
However, other signal components, such as
emission associated with SASI activity
\citep{Kuroda_16,Andresen_17} or additional PNS oscillation modes
\citep{Jakobus_23}, are also predicted in some models. {A more
thorough assessment of template-based searches for the signal
component studied in this work would also benefit from including
additional signal predictions. The models considered here represent
individual realisations of the underlying time-frequency structure,
and a larger sample would allow the bank's performance to be
characterised better.}
\end{enumerate}

Understanding how well different signal components
can be detected and reconstructed is important for determining
what physical information could be extracted from a future
core-collapse supernova GW detection. With only a limited number of
numerical waveform predictions currently available, the ability to
generate synthetic signals with \textsc{SynthGrav} provides a useful
tool for exploring the detectability of a wide range of possible
supernova GW signals.

\section*{Acknowledgements}
We thank Michele Zanolin for valuable feedback on the manuscript.
This work was enabled by resources provided by the National Academic Infrastructure for Supercomputing in Sweden (NAISS) at NSC partially funded by the Swedish Research Council through grant agreements no. 2022-06725 and no. 2018-05973. This work is supported by the Swedish Research Council (Project No. 2020-00452).

This material is based
upon work supported by the National Science Foundation Graduate Research Fellowship Program under Grant No.
2137424. Any opinions, findings, and conclusions or recommendations expressed in this material are those of the
authors and do not necessarily reflect the views of the National Science Foundation. Support was also provided by
the Graduate School and the Office of the Vice Chancellor for Research at the University of Wisconsin-Madison
with funding from the Wisconsin Alumni Research Foundation.

This research has made use of data or software obtained from the Gravitational Wave Open Science Center (gwosc.org), a service of the LIGO Scientific Collaboration, the Virgo Collaboration, and KAGRA. This material is based upon work supported by NSF's LIGO Laboratory which is a major facility fully funded by the National Science Foundation, as well as the Science and Technology Facilities Council (STFC) of the United Kingdom, the Max-Planck-Society (MPS), and the State of Niedersachsen/Germany for support of the construction of Advanced LIGO and construction and operation of the GEO600 detector. Additional support for Advanced LIGO was provided by the Australian Research Council. Virgo is funded, through the European Gravitational Observatory (EGO), by the French Centre National de Recherche Scientifique (CNRS), the Italian Istituto Nazionale di Fisica Nucleare (INFN) and the Dutch Nikhef, with contributions by institutions from Belgium, Germany, Greece, Hungary, Ireland, Japan, Monaco, Poland, Portugal, Spain. KAGRA is supported by Ministry of Education, Culture, Sports, Science and Technology (MEXT), Japan Society for the Promotion of Science (JSPS) in Japan; National Research Foundation (NRF) and Ministry of Science and ICT (MSIT) in Korea; Academia Sinica (AS) and National Science and Technology Council (NSTC) in Taiwan.

\section*{Data Availability}
The data used in this study can be obtained from 
the Gravitational Wave Open
Science Center. The signal we injected into the noise is publicly available. We will provide our Python scripts upon reasonable request.



\bibliographystyle{iopart-num}

\bibliography{sources}
\end{document}